\documentclass{PoS}

\title{Tantalizing dilaton tests from a near-conformal EFT}

\ShortTitle{Tantalizing dilaton tests from a near-conformal EFT      \hskip 1.23in   {\rm Julius Kuti and}}

\author{Zoltan Fodor\\
	University of Wuppertal, Department of Physics, Wuppertal D-42097, Germany\\
	Juelich Supercomputing Center, Forschungszentrum Juelich, Juelich D-52425, Germany\\
	Eotvos University, Pazmany Peter setany 1, 1117 Budapest, Hungary\\
	\email{fodor@bodri.elte.hu}}

\author{Kieran Holland\\
	University of the Pacific, 3601 Pacific Ave, Stockton CA 95211, USA\\
	Albert Einstein Center for Fundamental Physics, Bern University, Bern, Switzerland\\
	\email{kholland@pacific.edu}}

\author{\speaker{Julius Kuti}\\
	University of California, San Diego, 9500 Gilman Drive, La Jolla, CA 92093, USA\\
	\email{jkuti@ucsd.edu}}

\author{\speaker{Chik Him Wong}\\
	University of Wuppertal, Department of Physics, Wuppertal D-42097, Germany\\
	\email{cwong@uni-wuppertal.de}}

\abstract{
	
The dilaton low-energy effective field theory (EFT) of an emergent light scalar is probed  in the paradigm of strongly coupled near-conformal gauge theories. These studies are motivated	by models which exhibit  small  $\beta$-functions near the conformal window (CW), perhaps with  slow scale-dependent walking and a light scalar with ${ 0^{++} }$ quantum numbers. We report our results from the hypothesis of a dilaton inspired EFT analysis with two massless fermions in the two-index symmetric (sextet) representation of the SU(3) color gauge group. With important caveats in our conclusions, conformal symmetry breaking entangled with chiral symmetry breaking would drive the near-conformal infrared behavior of the theory predicting characteristic dilaton signatures 
of the light scalar from broken scale invariance when probed on relevant scales of fermion mass deformations. From a recently reasoned choice of the dilaton potential in the EFT description~\cite{Golterman:2016lsd} we find an unexpectedly light dilaton mass in the chiral limit at $m_d/f_\pi = 1.56(28)$, set in units of the pion decay constant $f_\pi$. Subject to further statistical and systematic tests of continued post-conference analysis, this result is significantly lower than our earlier estimates from less controlled extrapolations of the light scalar (the $\sigma$-particle) to the massless fermion limit of chiral perturbation theory. We also discuss important distinctions between the dilaton EFT analysis and the linear $\sigma$-model without dilaton signatures. For comparative reasons, we comment on dilaton tests from recent work with fermions in the fundamental representation with $n_f=8$ flavors.
}

\FullConference{The 36th Annual International Symposium on Lattice Field Theory - LATTICE2018\\
		22-28 July, 2018\\
		Michigan State University, East Lansing, Michigan, USA.}

\usepackage{graphicx}
\usepackage{comment}
\usepackage{wrapfig}
\usepackage{cite}
\usepackage{subcaption}
\usepackage{amsmath}
\usepackage{setspace}
\usepackage{ragged2e}
\usepackage{grffile}
\usepackage{cleveref}

\begin{document}

\section{Introduction and overview} \label{section:intro} 
\hspace{10pt}\parbox{5.3in}{\scriptsize\raggedright Tantalus, a king of ancient Phrygia in Greek mythology, made the mistake of gravely offending the gods. 
	As a punishment, once dead the king was forced to stand in a pool of water, with fruit hanging just over his head. 
    The water would recede every time the king tried to take a sip, and the fruit would lift away every time he reached to take a bite.}
\vskip 0.1in
New results are reported from the dilaton effective field theory analysis of a very light
scalar with  $0^{++}$  quantum numbers in lattice simulations of a  strongly coupled gauge theory, 
defined in the two-index symmetric fermion representation 
of the SU(3) color gauge group (one flavor doublet of massless fermions with sextet color). The sextet model thus defined
plays a prominent role in the near-conformal gauge theory paradigm, perhaps with
BSM implications. 

In earlier work we discovered the light $0^{++}$ scalar
as one of the most significant theoretical and practical consequences of near-conformal infrared behavior 
in the sextet theory~\cite{Fodor:2014pqa,Kuti:2014epa,Fodor:2016pls,Fodor:2016zil,Fodor:2017nlp}, 
radically different from the heavy $\sigma$-particle of 
gauge theories far from the conformal window and modeled like QCD. 
We investigate the hypothesis of a dilaton inspired EFT  in the sextet model with important caveats in our conclusions.
Accordingly,
conformal symmetry breaking would be entangled with chiral symmetry breaking $(\chi SB$) driving near-conformal 
infrared behavior and predicting characteristic dilaton signatures of the light scalar from broken scale invariance
when probed on relevant scales of fermion mass deformations. 
We find an unexpectedly light dilaton mass in the chiral limit at $m_d/f_\pi = 1.56(28)$, set
in 	units of the pion decay constant $f_\pi$ from a recently reasoned choice 
of the dilaton potential in the Lagrangian of the EFT~\cite{Golterman:2016lsd}.
Subject to further statistical and systematic  tests of continued post-conference analysis, this result is 
significantly lower than our earlier 
estimates for the $\sigma$-particle from less controlled extrapolations
to the massless fermion limit of chiral perturbation theory without characteristic dilaton features.

Two models are identified in Section~\ref{section:beta} with significantly 
smaller  step $\beta$-functions than QCD and perhaps 
with slowly walking scale dependence correlated with dilaton signatures of emergent light scalars.
Before pivoting to the dilaton analysis in Section~\ref{section:dilaton} we present first in Section~\ref{section:su2}  the unresolved 
challenges of the standard $\chi PT$ analysis and its extensions to the linear $\sigma$-model.
We also discuss important differences of the sextet dilaton analysis from chiral perturbation theory ($\chi PT$) and from general extensions of 
the linear $\sigma$-model without dilaton signatures.
In Section~\ref{section:dilaton} the dilaton EFT is discussed and important  predictions are reviewed for hypothesis testing.
Results are reported from the current status of the sextet dilaton analysis. 
Based on the recently published data of the LSD collaboration~\cite{Appelquist:2018yqe}
some of our own comparative $n_f=8$  dilaton analysis is also reported.
New ideas and simulations are briefly presented in Section~\ref{section:rmt} for reaching much reduced fermion mass scales in the 
$\epsilon$-regime of the dilaton EFT. We conclude in Section 6 with some cautionary remarks and caveats for the outlook.
\vskip -0.3in
\section{Near-conformal $\beta$-functions and the light ${\mathbf 0^{++} }$ scalar }
\label{section:beta}
\begin{figure}[h]
	
	\begin{subfigure}{0.65\textwidth}
		\includegraphics[width=0.98\linewidth, height=7.5cm]{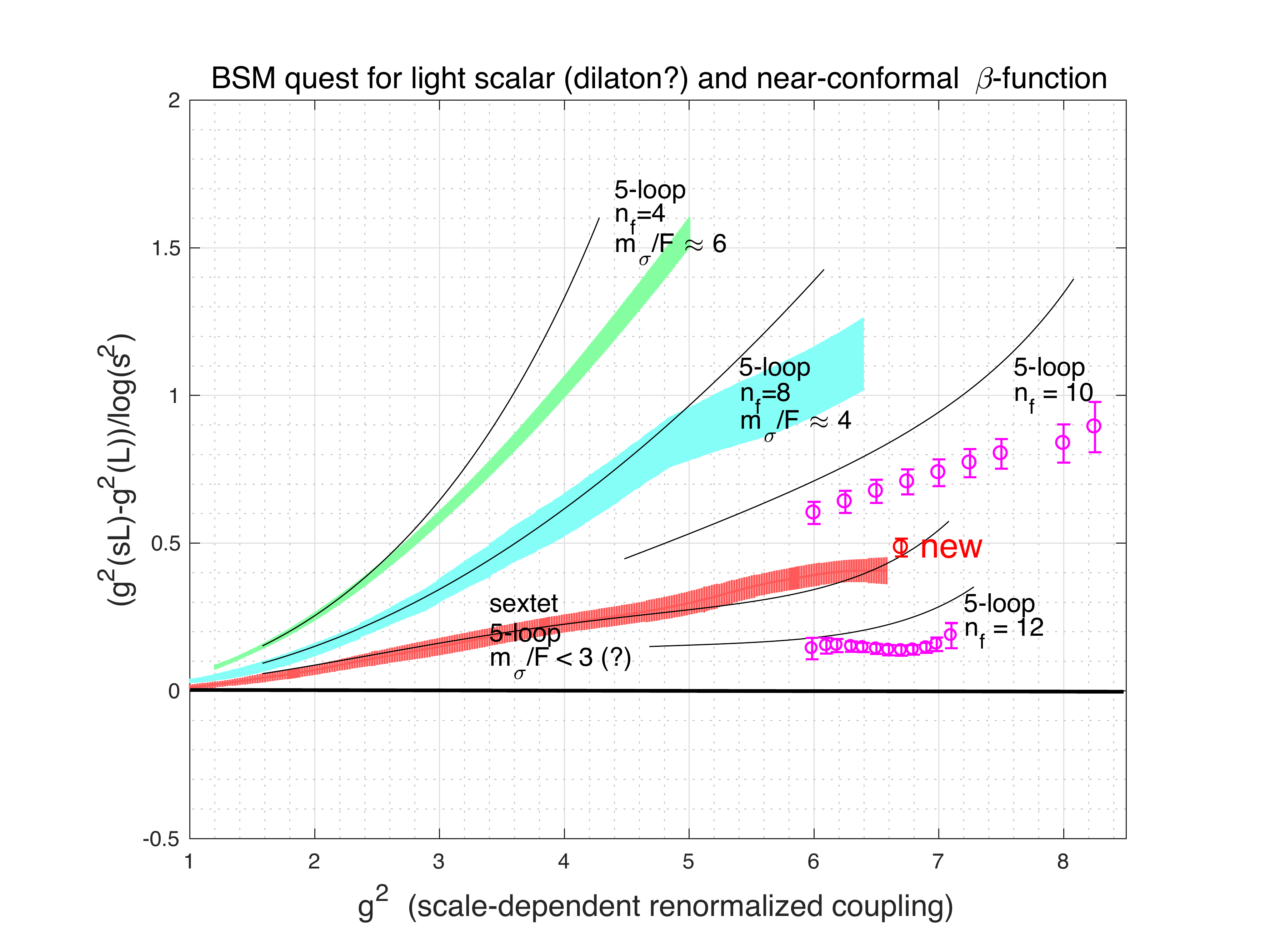} 
		\label{fig:subim1}
	\end{subfigure}
	\begin{subfigure}{0.38\textwidth}
		\includegraphics[width=0.9\linewidth, height=3cm]{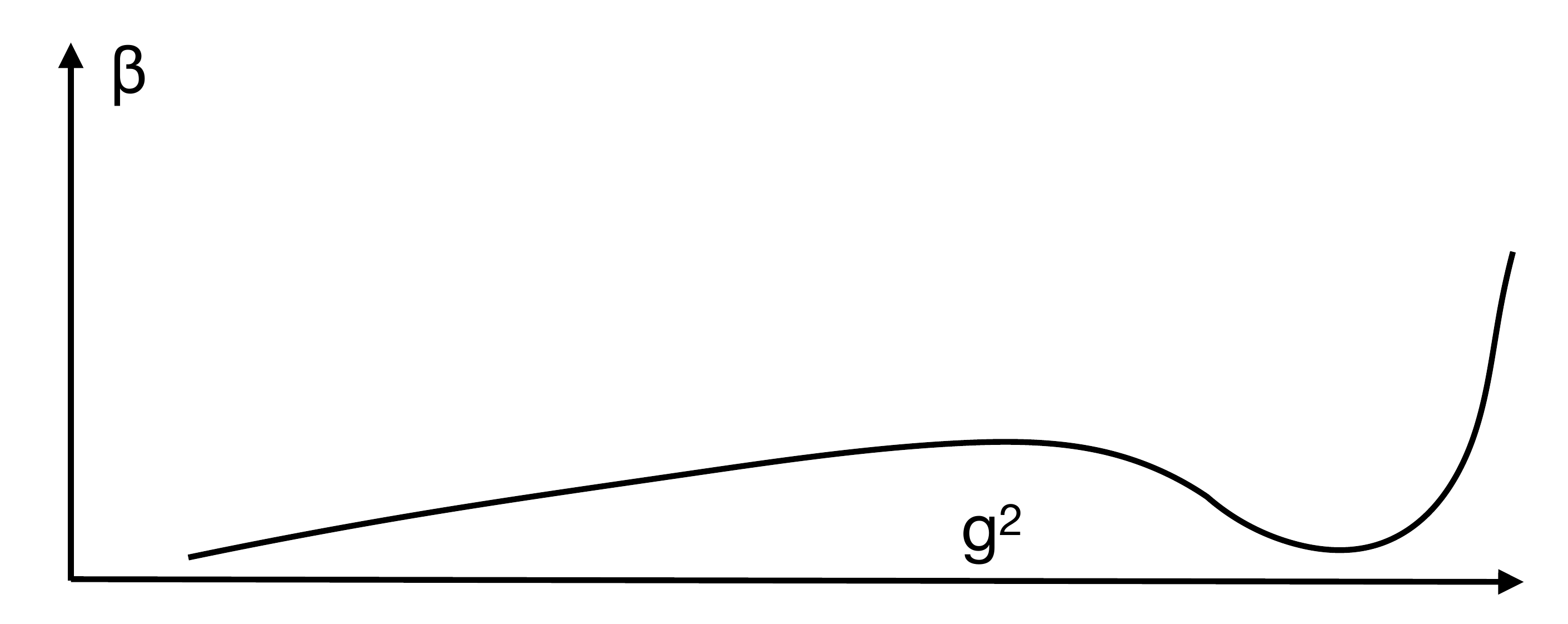}
		\label{fig:subim2}
	\end{subfigure}
	\vskip -0.1in
	\caption{\footnotesize The five $\beta$-functions of the models surveyed in this section are shown on the left panel. The correlation 
	with the emergent light scalars in the $n_f=8$ model and the sextet model are discussed in the text. The right panel is the cartoon version 
    of a walking $\beta$-function, replaced by small and flat $\beta$-functions in realistic near-conformal models.}
	\label{fig:beta5loop}
	\vskip -0.2in
\end{figure}

\vskip -0.1in
The step $\beta$-function of the renormalized gauge coupling  is shown from lattice studies in Fig.~\ref{fig:beta5loop} for two 
different fermion representations of strongly coupled gauge theories with SU(3) gauge group~\cite{Fodor:2017nlp}.
Five $\beta$-functions are shown and three of them may exhibit dilaton-like features of their light scalar.
In the fundamental representation, the $n_f=4$ model in Fig.~\ref{fig:beta5loop} is like QCD with four massless fermions.
It has the largest $\beta$-function~\cite{Fodor:2012td}  and a  $\sigma$-particle with a heavier mass, 
$m_\sigma/F \sim 6$, set by the scale 
of the Goldstone decay constant $F$ in the massless fermion limit of $\chi SB$~\cite{Appelquist:2018yqe}.
(For convenience, we will change the notation  to $F\equiv f_\pi$ in all other sections.)
At increased flavor number $n_f=8$ the $\beta$-function becomes reduced and noticeably 
more flat compared to QCD ~\cite{Fodor:2015baa} and exhibiting 
a scalar mass with significantly reduced value~\cite{Aoki:2014oha,Appelquist:2016viq,Appelquist:2018yqe}. 
For the observed range of fermion mass deformations at $n_f=8$, the scalar mass and the pion mass are tracking each other.
This is quite distinct from QCD, with motivation for the  $n_f=8$ dilaton analysis in~\cite{Appelquist:2017wcg,Appelquist:2017vyy}.
At $n_f=10$ the further increase of the flavor number leads to further reduction of the $\beta$-function~\cite{Fodor:2017gtj,Nogradi}
with some  unresolved controversy.
An infrared fixed point was reported in~\cite{Chiu:2017kza} with vanishing $\beta$-function at $g^2\sim 7 $
and reconsidered in follow-up work with improved systematics which left 
the conformal or near-conformal behavior of the $\beta$-function unresolved 
at strong gauge couplings~\cite{Chiu:2018edw}.
The model with the largest flavor number in the fundamental representation at $n_f=12$ is shown in Fig.~\ref{fig:beta5loop} with the 
smallest $\beta$-function exhibiting very flat dependence at strong coupling~\cite{Fodor:2017gtj,Fodor:2017nlp}.
We will return in the future to the interesting challenge this model presents very close to the CW,
perhaps near-conformal and walking but 
with  controversies from recent lattice work~\cite{Cheng:2014jba,Hasenfratz:2016dou} and from a new
conference contribution~\cite{Hasenfratz:2018wpq} suggesting conformal behavior.

The sextet $\beta$-function in Fig.~\ref{fig:beta5loop} is from~\cite{Fodor:2015zna} with the $m_\sigma/F$ ratio 
taken from~\cite{Fodor:2016pls}. The point marked as new was our Lattice 2017 conference 
contribution~\cite{Fodor:2017die}, bridging
the volume dependent step $\beta$-function and the scale dependent $\beta$-function of the 
p-regime in the infinite volume limit. 
We have now a set of new gauge ensembles in the sextet model to  extend the 
small and flat step $\beta$-function of  Fig.~\ref{fig:beta5loop}  toward stronger gauge couplings. 
The mark (?) next to the sextet model ratio $m_\sigma/F < 3$ 
is our indicator that the final ratio in the chiral limit requires further analysis, with this report contributing new results.

The above brief survey presents motivations for near-conformal tests of the $n_f=8$ model and the $n_f=2$ sextet model
for dilaton signatures. 
The right panel of Fig.~\ref{fig:beta5loop} is the cartoon version 
of a walking $\beta$-function, presumably replaced by small and flat $\beta$-functions in realistic near-conformal models.
Although the unrealistic cartoon shape of the $\beta$-function might be sufficient for dilaton analysis
when the conformal limit is approached in some parametric expansion, 
we remain unconvinced that it is a required feature of walking. 
Recent 5-loop results are only plotted in Fig.~\ref{fig:beta5loop} to indicate the state of the art in the perturbative loop expansion~\cite{Baikov:2016tgj}, 
perhaps for future theoretical analysis of walking, based on speculations for a pair of complex conformal fixed points 
of walking non-perturbative $\beta$-functions below the CW.
It did not escape our attention that investigations of this scenario should also include the $n_f=12$ model. 
%
\section{Challenges of the sextet $\chi$PT analysis and its linear $\sigma$-model extensions}\label{section:su2}

\subsection{Early discovery of the light scalar and the associated particle spectrum}
\begin{figure}[htb!]
	\begin{center}
		\begin{tabular}{cc}
			\includegraphics[width=0.45\textwidth]{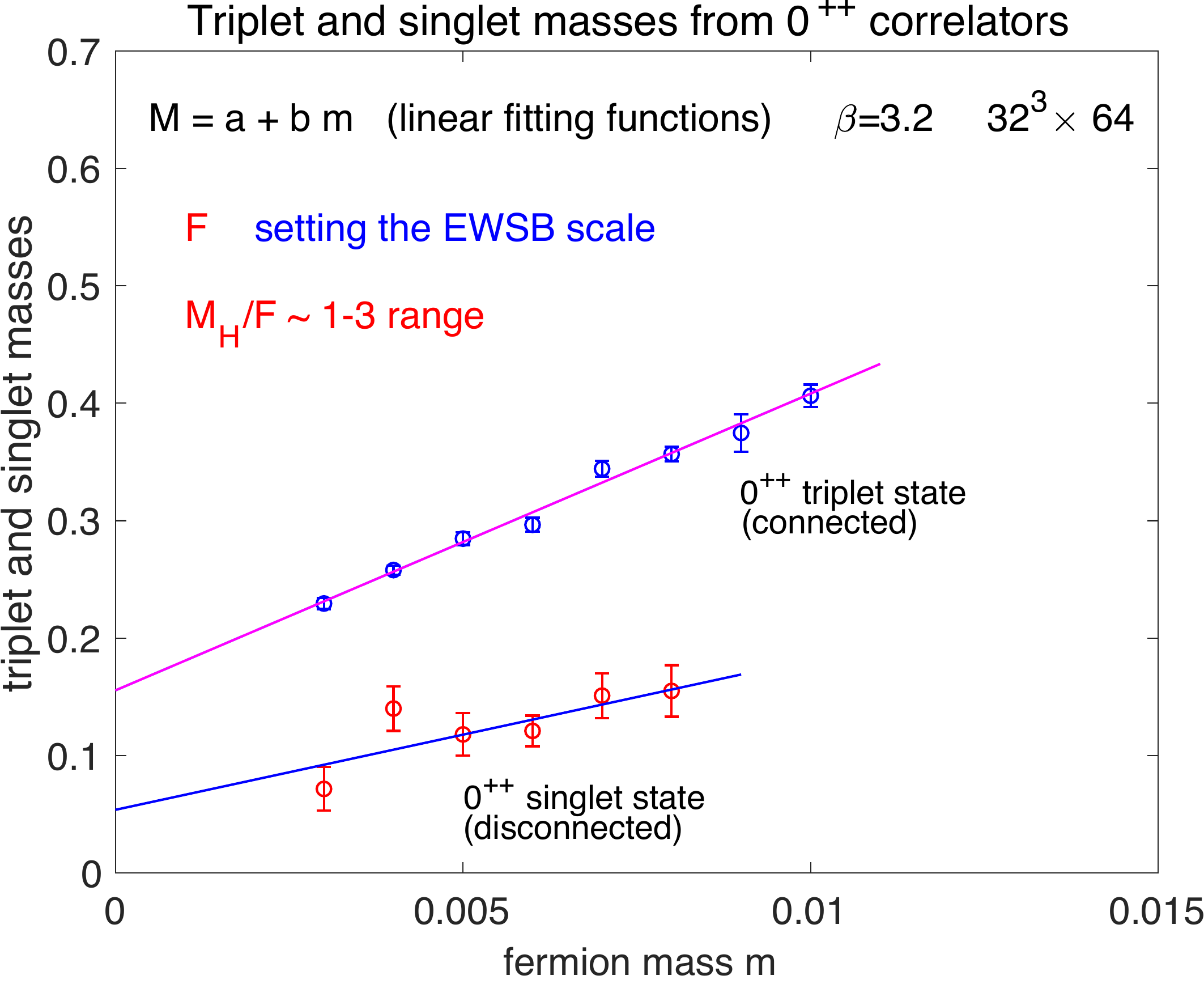}&
			\includegraphics[width=0.52\textwidth]{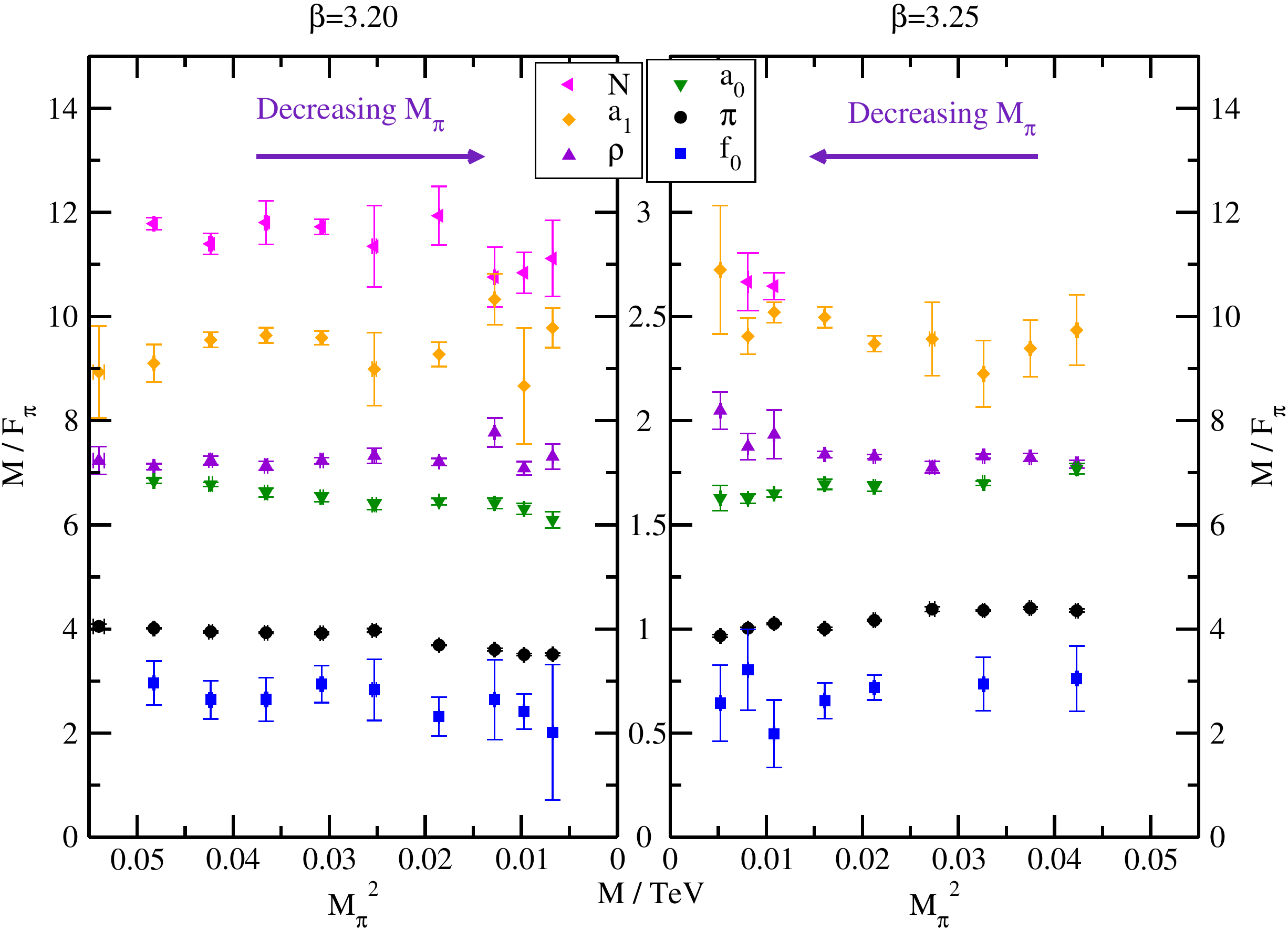}
		\end{tabular}
	\end{center}
	\caption{\footnotesize (left) The first result on the light sextet scalar from ~\cite{Fodor:2014pqa,Kuti:2014epa}; (right) The status of hadron spectroscopy
	in the sextet model is shown as reported in~\cite{Fodor:2016pls} including $M_\pi, F_\pi$ and the scalar mass $M_d$  
	($M_d\equiv M_H(f_0)$ in earlier notation).  The $M_\pi, F_\pi,M_d$ 
	data set of the analysis has been updated and refined since using new lattice ensembles in large volumes. Input data  $M_d$, as used in 
	some parts of the dilaton analysis, comes from the report in~\cite{Fodor:2016pls} as graphically represented on the right panel of the figure.}
	\label{fig:scalar}
\end{figure}

The light  $0^{++}$ scalar in the two-index symmetric (sextet) fermion representation of the SU(3) color gauge group was reported first at Lattice 2013 in ~\cite{Fodor:2014pqa,Kuti:2014epa} as shown in the left panel of Fig.~\ref{fig:scalar}.
We have our $M_\pi,F_\pi,M_d$ sextet data set  ($M_d\equiv M_H(f_0)$ in earlier notation) from a very large number of gauge ensembles at three lattice spacings in a range 
of fermion masses, $m=0.0010-0.0080$, 
with lattices sizes from $32^3\times 64$ to $64^3\times 96$. 
The finite size analysis of the data set was presented in~\cite{Fodor:2017nlp}. We use infinite volume extrapolations of  $M_\pi,F_\pi$ data sets
at fixed bare gauge coupling $\beta = 6/g^2$, with $\beta=3.20$ at each of the lowest five input fermion masses applied to the analysis. The $M_d$ input
is always taken from the largest volume of the gauge ensembles at each input fermion mass.

\subsection{Pivot to dilaton EFT from ${\boldmath \chi}$PT and its linear sigma model extensions}
\begin{figure}[ht]
	\centering
	\parbox{.62\textwidth}{
		\begin{subfigure}{.49\linewidth}		
			\caption*{\footnotesize $\chi PT$ fit to $M^2_\pi$:}		
			\includegraphics[width=\textwidth]{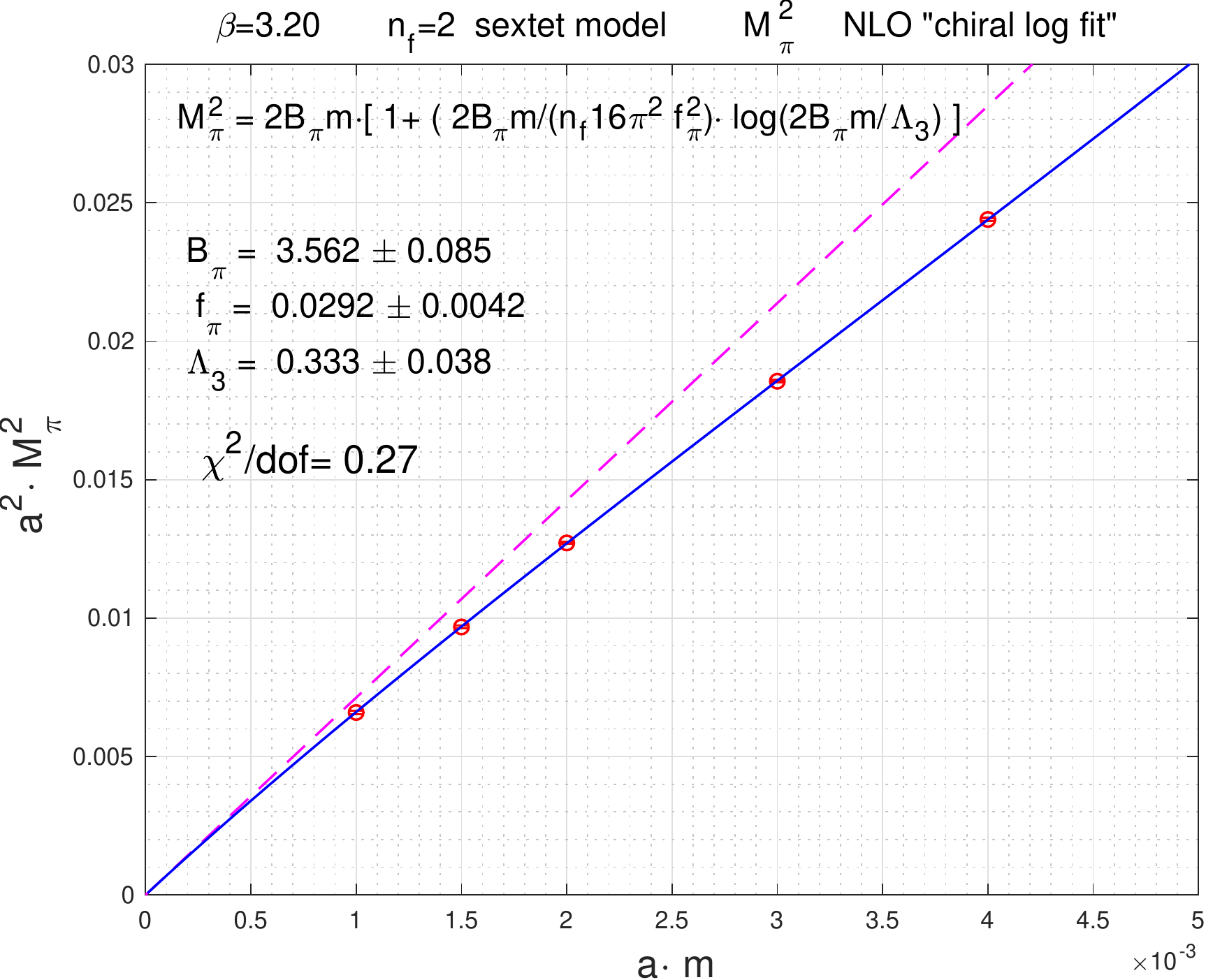}	
		\end{subfigure}
		\begin{subfigure}{.49\linewidth}
				\caption*{\footnotesize $\chi PT$ fit to $F_\pi$:}			
			\includegraphics[width=\textwidth]{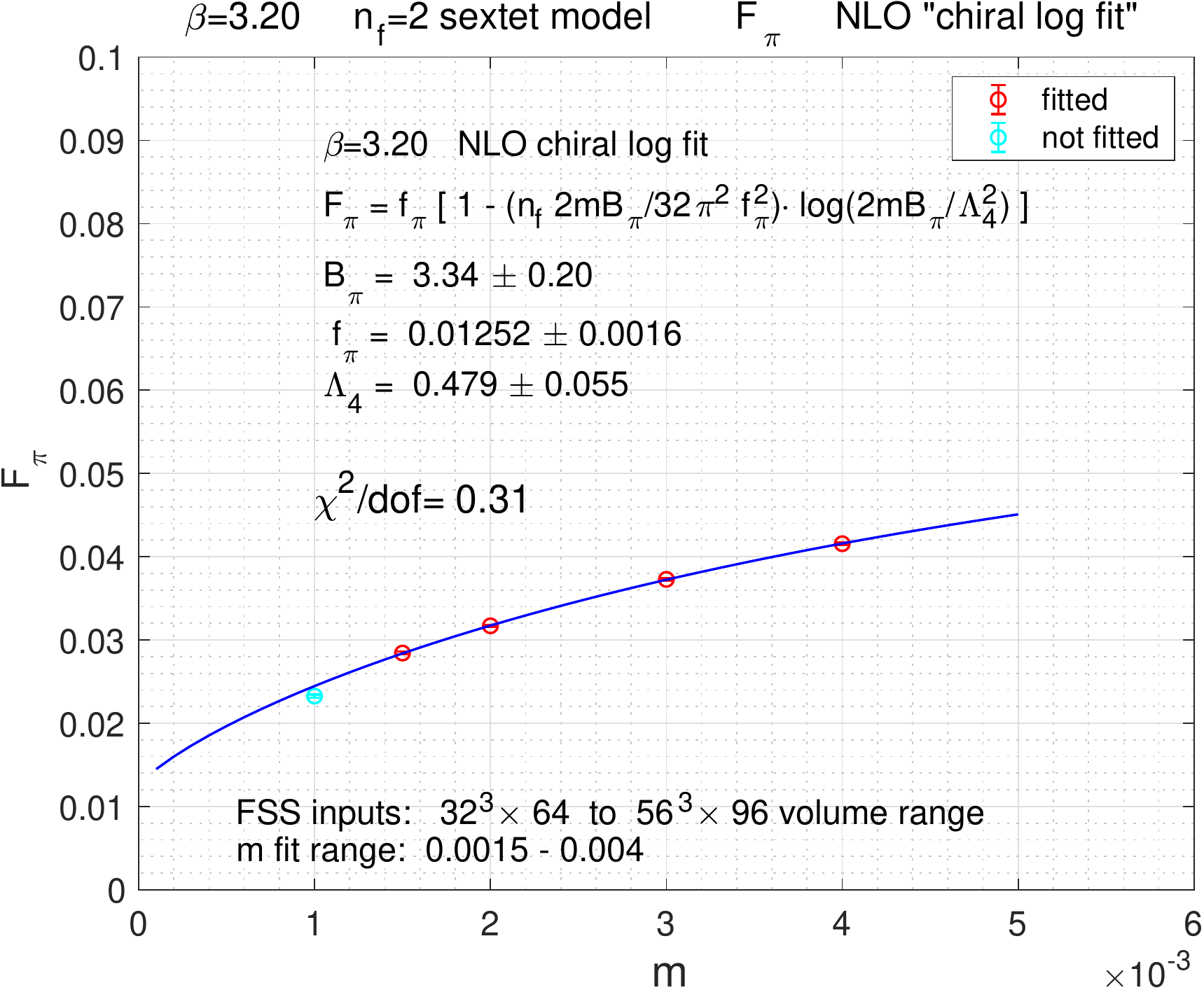}
		\end{subfigure}\\
		\begin{subfigure}{.49\linewidth}
           \caption*{\footnotesize Conformal fit to $M_\pi$:}			
			\includegraphics[width=\textwidth]{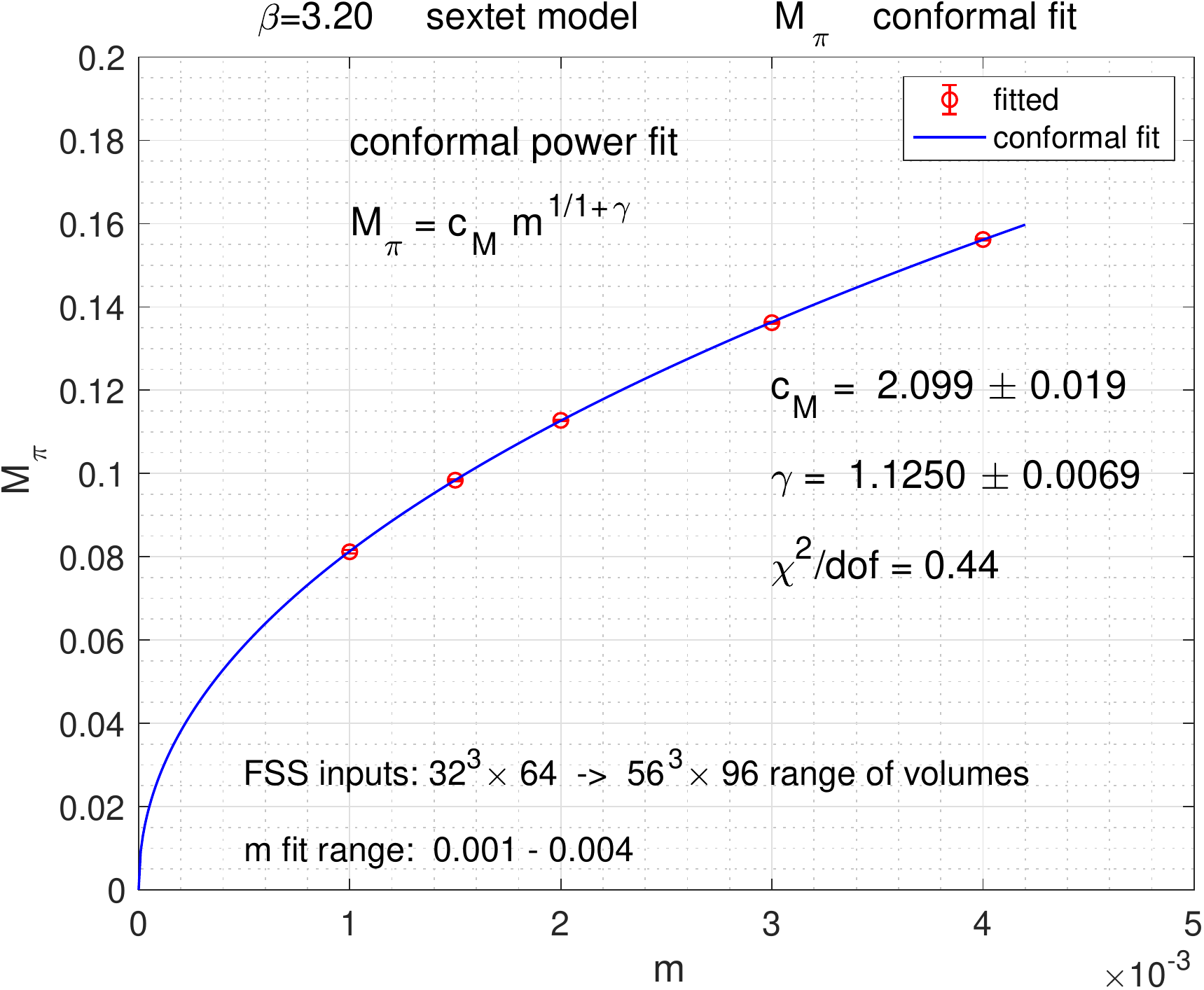}
		\end{subfigure}
		\begin{subfigure}{.49\linewidth}
			\caption*{\footnotesize Conformal fit to $F_\pi$:}			
			\includegraphics[width=\textwidth]{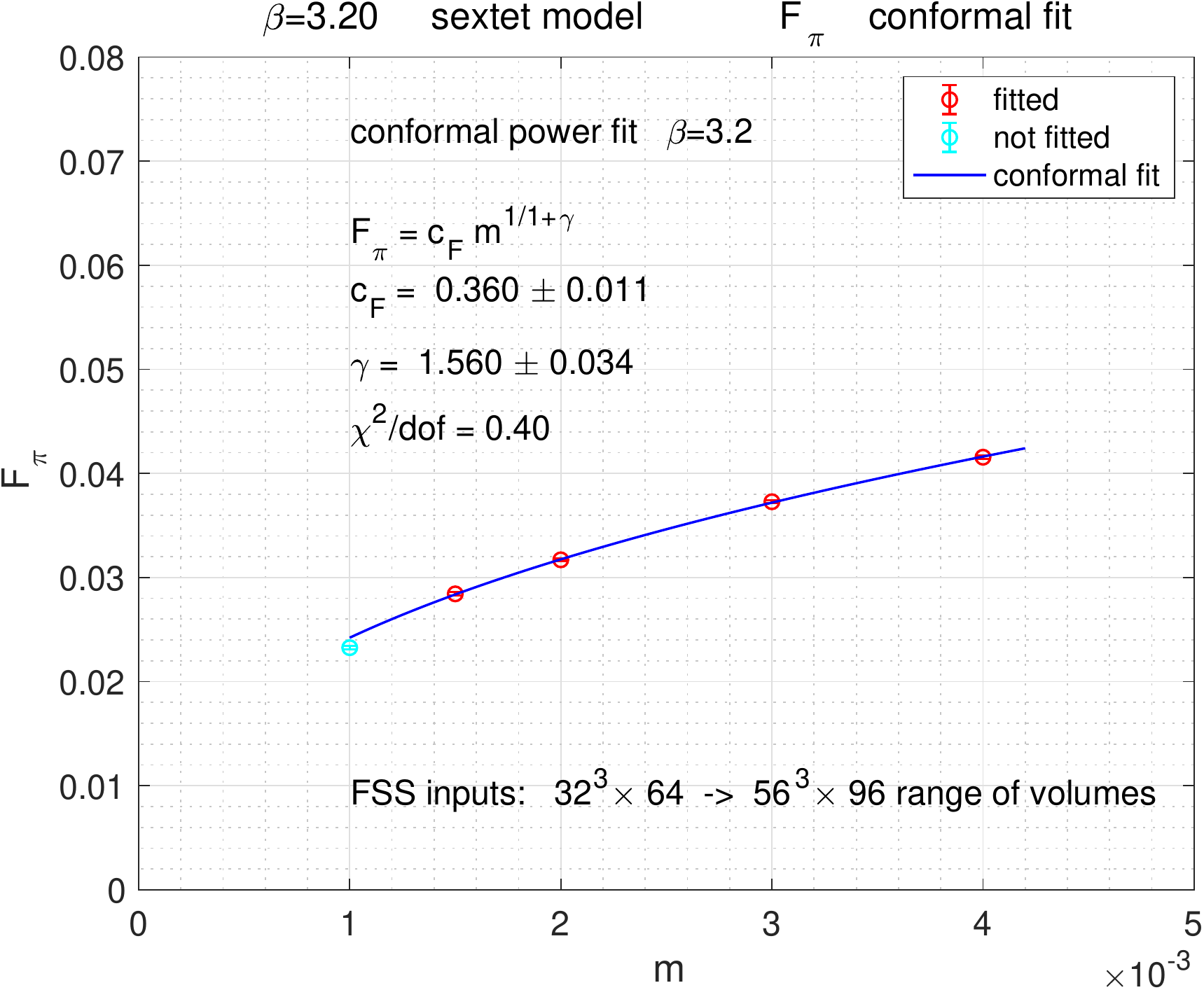}
		\end{subfigure}
}
	\begin{subfigure}{.33\textwidth}
		\includegraphics[width=\textwidth]{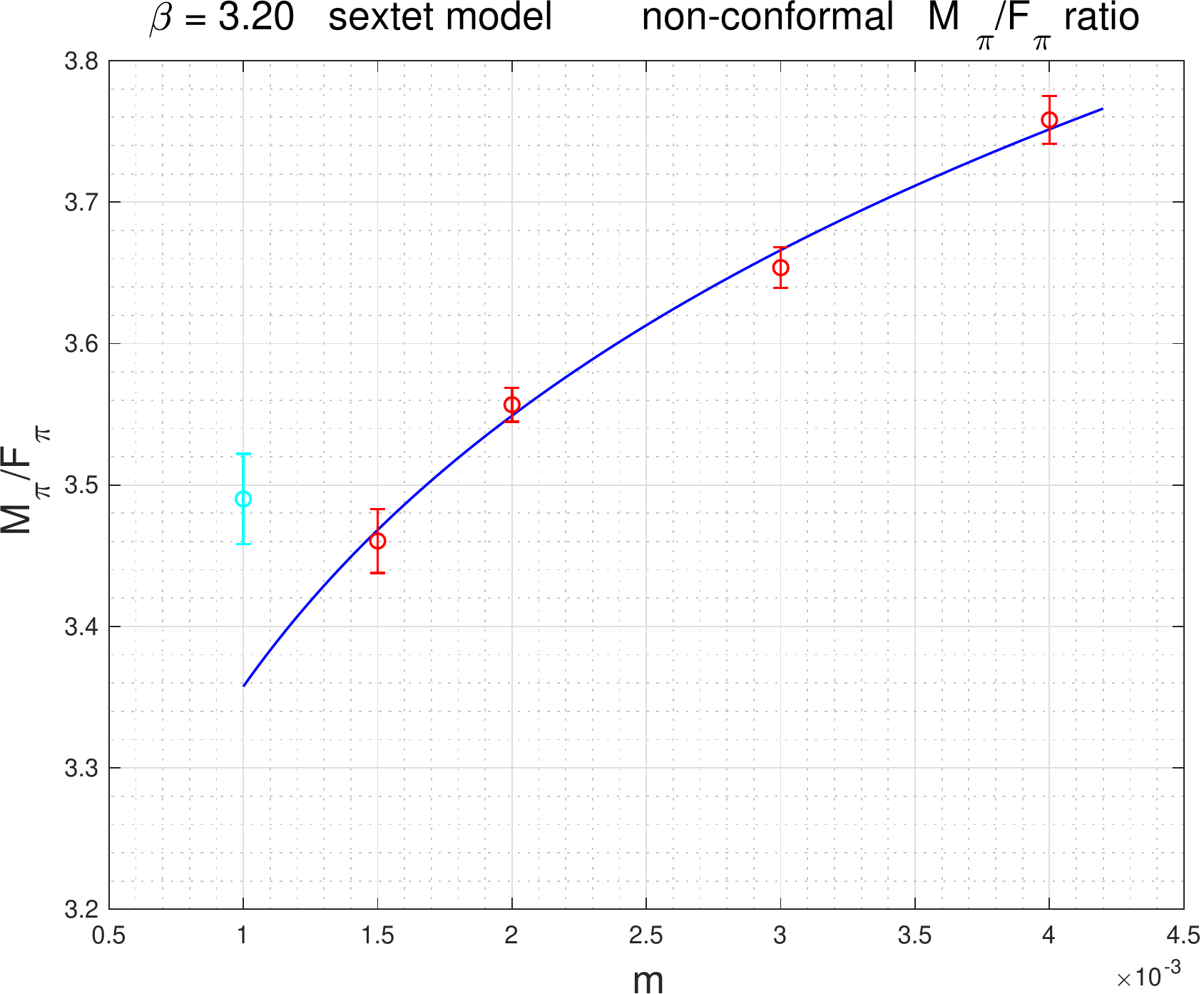}
		\caption*{\footnotesize Failed conformal fit with large variation of the  $M_\pi/F_\pi$ ratio 
		in the fitted fermion mass range.}
	\end{subfigure}\hfill
	\caption{\footnotesize  Chiral and conformal fits of the $M_\pi,F_\pi$ parameters in the sextet model.
	The lowest $m$ is not fitted in $F_\pi$ from incomplete finite size scaling analysis.}	
	\label{fig:rep6chiPT}
	\vskip -0.2in
\end{figure}

Motivated by the $SU(2)$ doublet of mass deformed Goldstone pions of the sextet model, we tested mass deformed chiral perturbation theory ($\chi PT$)
when applied to the above described sextet data for $M_\pi$ and $F_\pi$. As shown in Fig.~\ref{fig:rep6chiPT}, 
the logarithmic form of  NLO one-loop $\chi PT$ can 
be separately fitted to $M_\pi$  with the three parameters $B_\pi,f_\pi,\Lambda_3$ and good $\chi^2$ for the one-loop chiral Lagrangian. Similarly,  $F_\pi$ 
fits well with a separate set of three parameters $B_\pi,f_\pi,\Lambda_4$ and good $\chi^2$.
However, NLO $\chi PT$ fails for simultaneous fits of the $M_\pi,F_\pi$ input set because the two pairs $B,f_\pi$ of
low-energy parameters of the $\chi PT$ Lagrangian are inconsistent in separate fits of $M_\pi$ and $F_\pi$. 

Forcing conformal behavior on the sextet model is not the answer. 
The results in Fig.~\ref{fig:rep6chiPT}  from forced conformal fits to $M_\pi$ and $F_\pi$ show unacceptable results  in 
the mass range of our analysis with significant variation of the  $M_\pi/F_\pi$ ratio and inconsistent 
conformal $\gamma$ exponents, statistically rejected on any reasonable level of confidence. 
We remain puzzled and unconvinced by recent claims of observing conformal behavior in the model~\cite{Hansen:2017ejh}. 

As a possible remedy to the failures of $\chi PT$, we had been experimenting in the past with rooted staggered $\chi PT$ to include cutoff 
effects with taste breaking from the 
staggered pion spectrum. The results at $\beta=3.20$ were reported in~\cite{Fodor:2017nlp} but applying the same analysis at $\beta=3.25$ new 
problems emerged. The plausible interpretation of the failing $\chi PT$ analysis is the  light scalar closely tracking the 
mass-deformed pion spectrum as shown in Fig.~\ref{fig:scalar}. Pion dynamics remains closely coupled to $0^{++}$ scalar dynamics, violating
the basic premise of  $\chi PT$.

In the sextet model with  ${ SU(2)\times SU(2)}$ flavor group the linear $\sigma$-model would be a
natural candidate to extended ${ \chi PT}$ to the coupling of the Goldstone triplet to the  $\sigma$-particle in the ${ m_{\sigma}/F < 3}$ 
mass range with broken ${SU(2)\times SU(2)}$${\sim O(4)}$ symmetry.
This would match the construction of the standard model
Higgs sector at a higher ${ m_\sigma}$ mass  in the presence of 
fermion mass deformations, before extrapolation to the chiral limit is taken.  However, the linear $\sigma$-model with  
broken ${ SU(2)\times SU(2)\sim O(4)}$ symmetry is not applicable in the current range of fermion 
mass deformations where existing data with ${ m_{\sigma}\sim m_{\pi}}$ do not comply with the condition  ${ m^2_{\sigma}\geq 3m^2_{\pi}}$ 
derived from the tree-level Lagrangian of the linear $\sigma$-model~\cite{jkuti}. Much lower fermion mass deformations 
would be needed to probe  the linear $\sigma$-model regime. Following earlier work of Soto~\cite{Soto:2011ap},
 we were experimenting with extensions of the linear $\sigma$-model to find relevant additional terms for the analysis, similar to Eq.(\ref{eq:EFT}) with the choice $V_\sigma$ from Eq.(\ref{eq:Vsigma})
 but with preset value of the $y$ exponent at  $y=2$ in the mass term of the Lagrangian, making the dilaton EFT practically 
 indistinguishable from the generalized linear $\sigma$-model.
 We performed one-loop calculations at $y=2$ and made tree level fits to our sextet lattice ensembles without satisfactory results  
 ( in the meantime, complete one-loop  calculations appeared in~\cite{Hansen:2016fri}). 
 While working on loop correction in the analysis and the relaxation of the $y=2$ constraint to accommodate the more general dilaton analysis
 of the sextet model,  the pioneering tree level $n_f=8$ dilaton analysis appeared~\cite{Appelquist:2017wcg} 
 and the two groups were off to the dilaton races~\cite{Appelquist:2017wcg,Fodor:2017gtj}. In this spirited competition we published the first sextet analysis~\cite{Fodor:2017gtj}.
A follow-up publication to~\cite{Appelquist:2017wcg} appeared~\cite{Appelquist:2017vyy} extending the  $n_f=8$ analysis and including
some limited sextet analysis based on
plots from our earlier sextet publications without access to our exclusive and more recent data sets. 
In the next section we provide some technical details of our significantly extended dilaton analysis beyond what was published in~\cite{Fodor:2017gtj}
for the sextet model. For comments and comparative purposes, we also include our extended $n_f=8$ analysis based on recently published data from~\cite{Appelquist:2018yqe}.

\section{Dilaton effective field theory of two candidates for near-conformal models}\label{section:dilaton}
\subsection{The EFT Lagrangian}
In Section~\ref{section:beta} we presented evidence for the correlated trend between the reduced mass of 
the $0^{++}$ singlet scalar and the reduced  size of the $\beta$-function close to the lower edge of the CW 
suggesting near-conformal behavior in the sextet model.
In the currently accessible range  of fermion mass deformations,
the mass of the light scalar is tracking closely the Goldstone boson (pion) multiplet from spontaneous chiral symmetry
breaking (${ \chi SB}$) of the underlying ${ SU(n_f)\times SU(n_f)}$ flavor group, with $n_f=2$ in the sextet model. 
This characteristic behavior is captured in a recently investigated low-energy 
EFT~\cite{Golterman:2016cdd,Golterman:2016lsd,Appelquist:2017wcg,Appelquist:2017vyy} to describe the light 
$\sigma$-particle with $0^{++}$  quantum numbers, coupled to pion dynamics as a dilaton from broken scale invariance.
The first application to the $n_f=8$ model was reported in~\cite{Appelquist:2017wcg,Appelquist:2017vyy}. After our first 
analysis of the sextet model in~\cite{Fodor:2017nlp} we report here a broader scope of dilaton EFT tests in the sextet
and $n_f=8$ models. 

The minimal modification of the chiral Lagrangian with dilaton couplings leads to the EFT
\begin{equation}
{\cal L} = \frac{1}{2}\partial_{\mu} \chi \partial_{\mu} \chi\,-\,V(\chi) + \frac{f^2_\pi}{4}\big(\frac{\chi}{f_d}\big)^2 
~{\rm tr}\big[\partial_{\mu}\Sigma~\partial_{\mu}\Sigma^\dagger\big] +
\frac{m^2_\pi f^2_\pi}{4}\big(\frac{\chi}{f_d}\big)^y ~ {\rm tr}\big[\Sigma + \Sigma^\dagger\big],\label{eq:EFT}
\end{equation}
where the notation $\chi = f_d\cdot e^{\sigma/f_d}$  is introduced for the  connection between the dilaton field $\sigma(x)$ and 
the compensator field $\chi(x)$ which 
transforms as $\chi(x) \rightarrow \chi'(x') =e^\omega \chi(x)$ 
under the shift $\sigma(x) \rightarrow \sigma'(x')=\sigma(x) + \omega\cdot f_d$ for scale transformations $x_\mu \rightarrow x'_\mu = e^{-\omega}x_\mu$.
The notation $f_d$ designates the minimum of the dilaton potential $V(\chi)$ in the chiral limit of vanishing fermion masses.
In Eq.~(\ref{eq:EFT}) of the EFT two different forms of the dilaton potential were chosen for our analysis,
\begin{subequations}
	\begin{align}
	V(\chi) \rightarrow V_d(\chi)  &= \frac{m^2_d}{16f^2_d}~\chi^4\big(4~{\rm ln} \frac{\chi}{f_d} - 1\big),  \label{eq:Vd} \\
	V(\chi) \rightarrow V_\sigma(\chi) &= \frac{m^2_d}{8f^2_d}~(\chi^2-f_d^2)^2.
	\label{eq:Vsigma}
	\end{align}
\end{subequations}
Recent theoretical motivation of Eq.(\ref{eq:Vd}) originates from~\cite{Golterman:2016lsd}, 
based on a parametric expansion of $V(\chi)$ as the CW is approached.  Eq.(\ref{eq:Vsigma}) is 
a  linear $\sigma$-model  inspired dilaton in~\cite{Goldberger:2008zz,Appelquist:2017wcg,Appelquist:2017vyy}.
We do not dwell on various aspects of the two choices in our first extended sextet tests.
The primary focus is on comparing the two choices $V_\sigma,V_d$ and commenting on what was reported in earlier work
~\cite{Appelquist:2017wcg,Appelquist:2017vyy,Golterman:2018mfm} and at the conference~\cite{Golterman:2018bpc}.
We will show that the choice of $V_d$, favored by the theory in~\cite{Golterman:2016lsd}, might become a game changer in the 
interpretation of the sextet model.

The Goldstone pions in Eq.(\ref{eq:EFT}) are described by the unitary matrix field $\Sigma~=~{\rm exp}[2i\pi/f_\pi]$ 
where the pion field is represented as $\pi=\Sigma_a\pi^aT^a$ with $n_f^2-1$ generators of the $SU(n_f)$ flavor group. 
We keep the same notation as in~\cite{Appelquist:2017wcg,Appelquist:2017vyy} for the parameters in Eqs.(\ref{eq:EFT},\ref{eq:Vd},\ref{eq:Vsigma}) for the
convenience of easy 
comparison with our analysis. In this notation, the tree level pion mass would be $m^2_\pi = 2B_\pi m$ close to the chiral limit,  
with the dilaton decoupled from pion dynamics.
The pion decay constant $f_\pi$ is defined in the chiral limit. The tree-level dilaton mass in the chiral limit
of vanishing fermion mass is designated as $m_d$ and it is
defined by the second derivative of the tree-level dilaton potential at its $\chi = f_d$ minimum as $V''(\chi=f_d)=m_d^2$. The dilaton mass
at finite fermion mass deformations is designated by $M_d$.

The scale-dependent anomalous dimension of the chiral condensate, as $y=3-\gamma$  in Eq.(\ref{eq:EFT}),  will require some more refined 
scale setting definition in walking theories and will not be addressed here. In the sextet model we have detailed information on the 
scale-dependent $\gamma$ which will be compared with the results emerging from the analysis of Eqs.(\ref{eq:EFT},\ref{eq:Vd},\ref{eq:Vsigma}).

The Lagrangian of the dilaton EFT in Eq.(\ref{eq:EFT}) has a long history which includes~\cite{Migdal:1982jp,Ellis:1984jv,Bardeen:1985sm,Leung:1989hw,Donoghue:1991qv,
	Goldberger:2008zz,Chacko:2012sy,Vecchi:2010gj,Matsuzaki:2013eva,Aoki:2016wnc,Crewther:2013vea,Cata:2018wzl} with further references. 
\subsection{MCMC analysis of dilaton EFT predictions from ${\mathbf V_d}$ and ${\mathbf V_\sigma}$ potentials:  sextet model}
\label{section:rep6V1}
In Markov Chain Monte Carlo (MCMC) based analysis of the implicit Maximum Likelihood (IML) procedure,  the targeted five physical parameters 
$f_\pi, B_\pi, y, m_d/f_\pi,f_d/f_\pi$ are defined by tree-level application of the dilaton EFT, based on  Eq.(\ref{eq:EFT}). 
For the choice of the dilaton potential $V_d(\chi)$ the physical parameters are subject to three 
non-linear constraints at each input fermion mass $m$ leading to twelve constraints with input at four different fermion masses
in the IML procedure,
\begin{align} 
&M_\pi^2\cdot F_\pi^{2-y} - 2B_\pi\cdot f_\pi^{(2-y)}\cdot m=0, \label{eq:Cd}\\
&F_\pi^{(4-y)}\cdot{\rm log}(F_\pi/f_\pi) - y\cdot n_ff_\pi^{(6-y)}B_\pi\cdot m/m_d^2f_d^2=0, \label{eq:Ad}\\
%
%
&(F_\pi^2/M_\pi^2)\cdot(3{\rm log}(F_\pi/f_\pi)+1) - (M_d^2/m_d^2)\cdot (f_\pi^2/M_\pi^2) - y(y-1) n_ff_\pi^4/2m_d^2f_d^2 =0. \label{eq:Md}
\end{align} 
The general scaling relation of Eq.(\ref{eq:Cd}) is independent from the choice of the dilaton potential~\cite{Golterman:2016cdd,Appelquist:2017wcg}. 
The dilaton potential  $V_d$  leads to two added non-linear conditions, with Eq.(\ref{eq:Ad}) set by\linebreak
$V'_d(\chi=F_d)$, and Eq.(\ref{eq:Md}) 
set by $V''_d(\chi=F_d)$, as  in~\cite{Appelquist:2017wcg}.
With unchanged scaling relation from Eq.(\ref{eq:Cd}), two alternative equations are 
derived from $V'_\sigma(\chi=F_d)$ and $V''_\sigma(\chi=F_d)$,
\vskip -0.3in
\begin{align} 
%
&F_\pi^{(4-y)}\cdot(1-f_\pi^2/F_\pi^2) -  2y\cdot n_ff_\pi^{(6-y)}B_\pi/m_d^2f_d^2\cdot m = 0, \label{eq:A1}\\
%
%
&3F_\pi^2/M_\pi^2-f_\pi^2/M_\pi^2 - 2M_d^2/m_d^2\cdot f_\pi^2/M_\pi^2 - y(y-1) n_ff_\pi^4/m_d^2f_d^2 =0.\label{eq:Md1}
\end{align} 
%
%
\begin{figure}[h!]
	\centering
	\parbox{.53\textwidth}{
		\begin{subfigure}{.49\linewidth}
			\caption*{\footnotesize posterior $f_\pi$ for rep6  $V_d$:}
			\includegraphics[width=\textwidth]{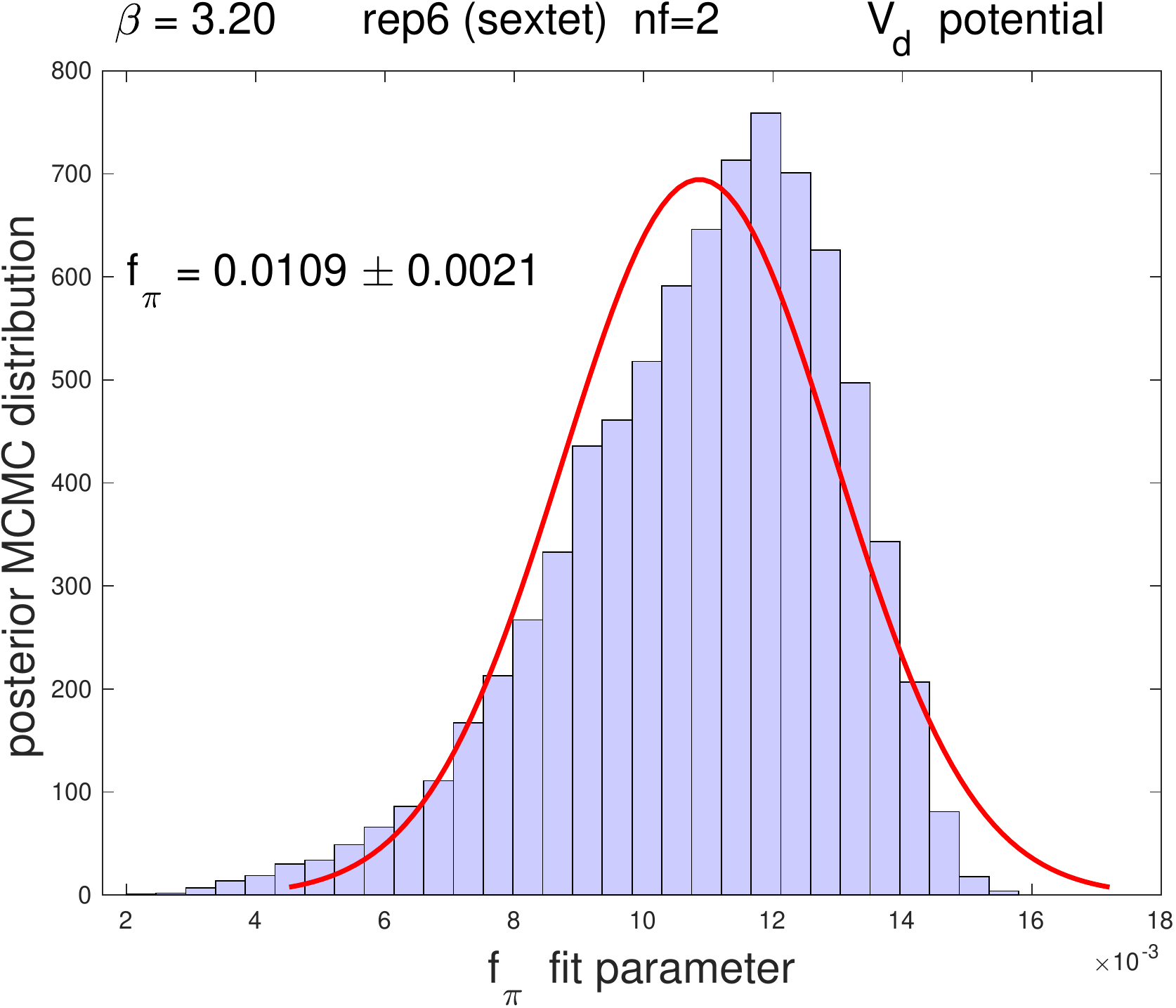}
		\end{subfigure}
		\begin{subfigure}{.49\linewidth}
			\caption*{\footnotesize posterior $B_\pi$ for rep6  $V_d$:}
			\includegraphics[width=\textwidth]{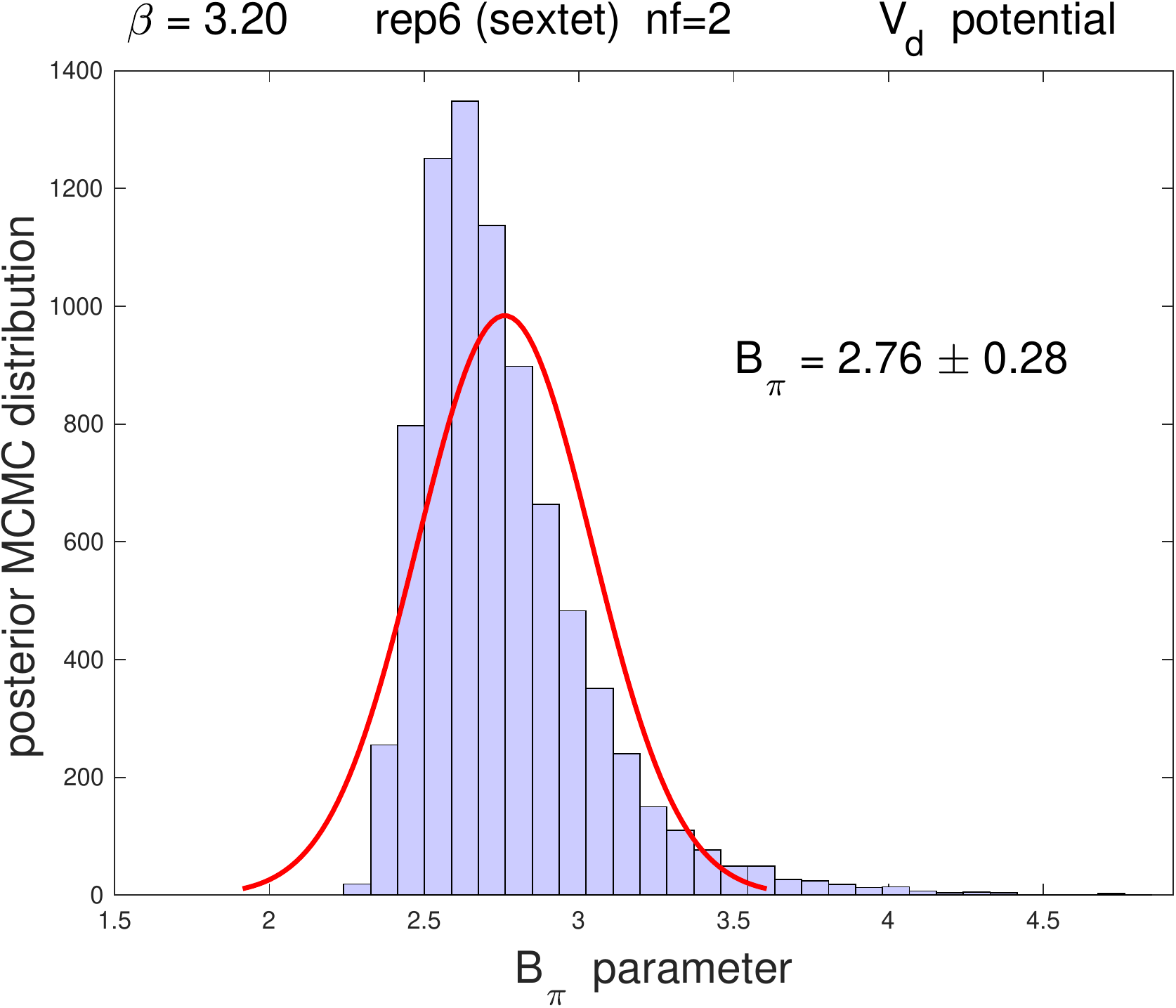}
		\end{subfigure}\\
		\begin{subfigure}{.49\linewidth}
			\caption*{\footnotesize posterior $\gamma$ for rep6  $V_d$:}
			\includegraphics[width=\textwidth]{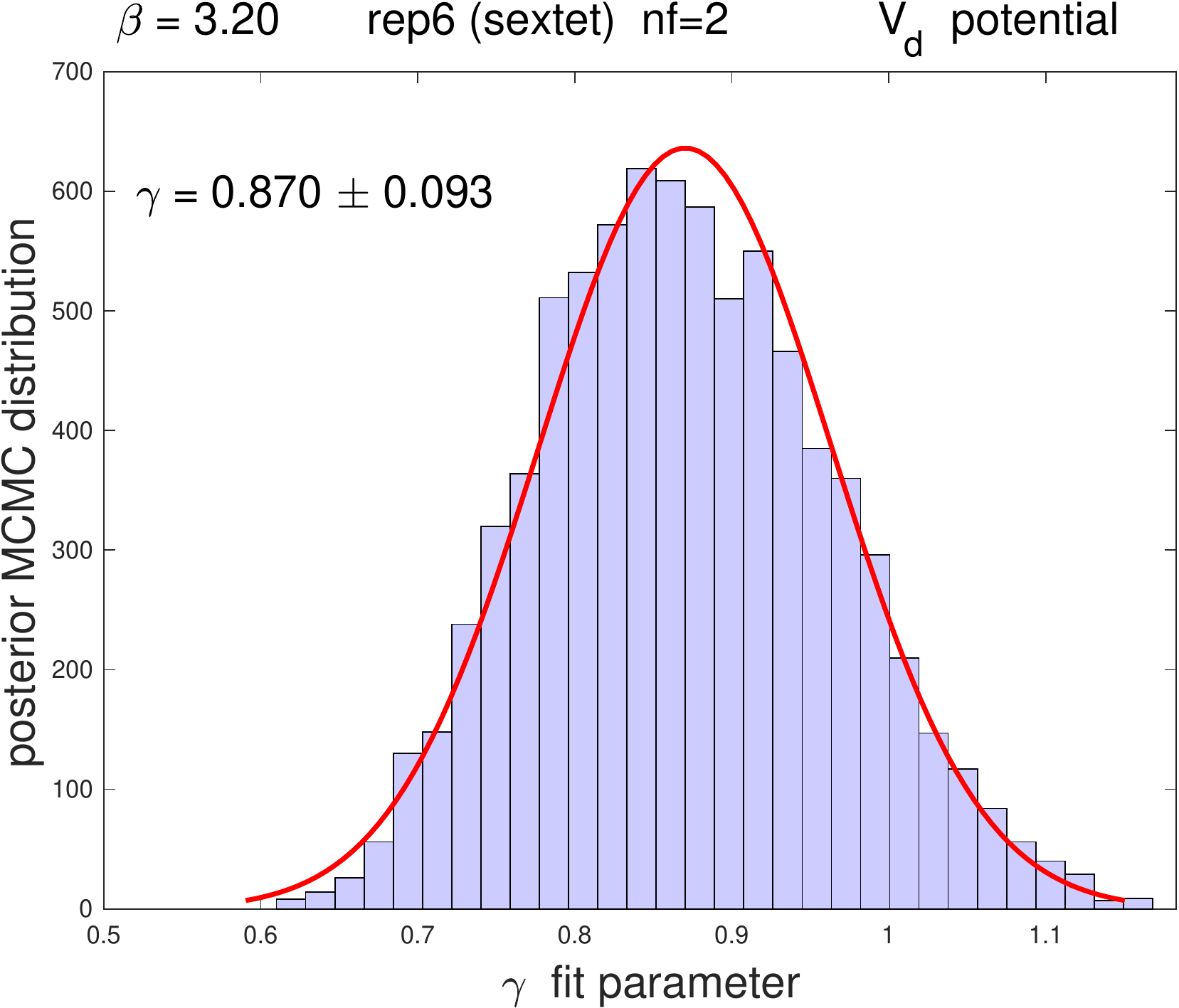}
		\end{subfigure}
		\begin{subfigure}{.49\linewidth}
			\caption*{\footnotesize posterior $m_d/f_\pi$ for rep6  $V_d$:}
			\includegraphics[width=\textwidth]{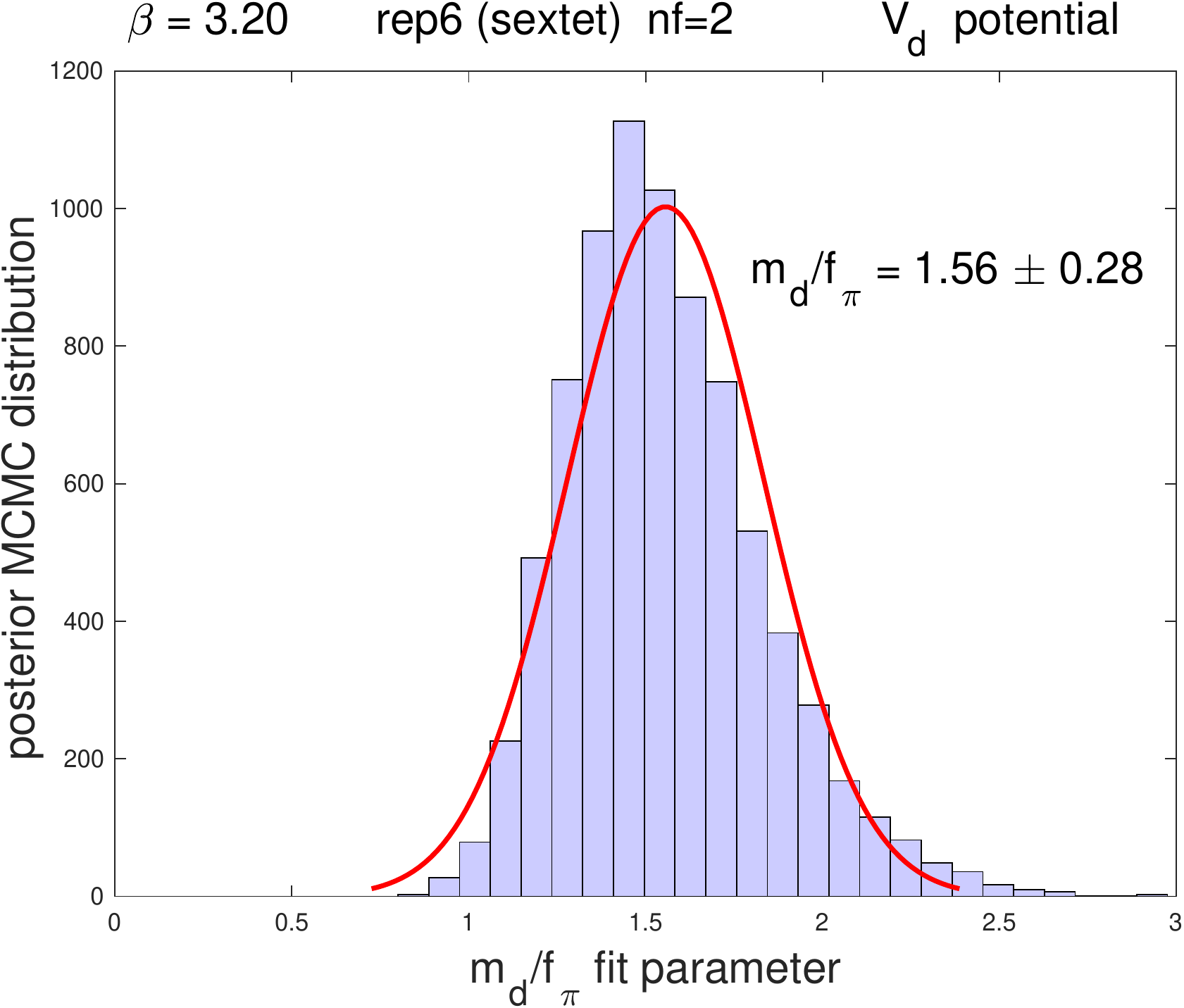}
		\end{subfigure}
	}
	\begin{subfigure}{.44\textwidth}
		\includegraphics[width=\textwidth]{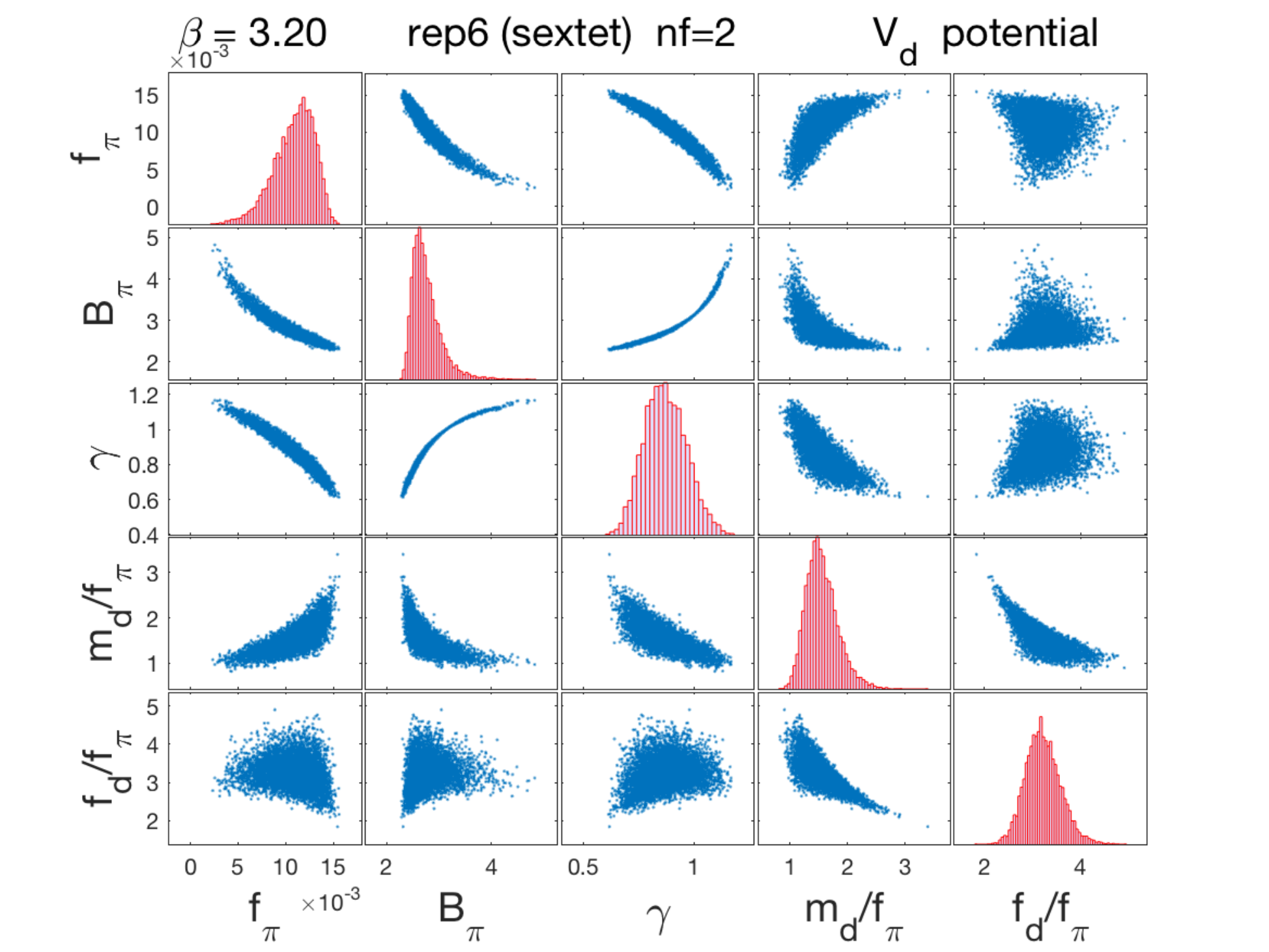}
		\caption*{\footnotesize Matrix plot of the five fitted physical parameters with posterior histogram in the diagonal 
			and off-diagonal scatter plots of the correlations. Four of the histograms are also shown on the left with fitted means and
		   1$\sigma$ equivalent percentile errors (the distributions are close to normal).
		}
	\end{subfigure}\hfill
	\caption{\footnotesize  MCMC based posterior probability distributions of five fitted physical parameters 
		and their correlations for the $V_d$ choice of the dilaton potential (rep6 sextet model). 
		The MCMC algorithm is explained in the text. Red lines indicate fits of normal distributions 
		to the histograms which show some deviations from the Gaussian shape as expected.
	}
	\label{fig:rep6Vd}
\end{figure}
\begin{figure}[h!]
	\centering
	\parbox{.53\textwidth}{
		\begin{subfigure}{.49\linewidth}
			\caption*{\footnotesize posterior $f_\pi$ for rep6  $V_\sigma$:}
			\includegraphics[width=\textwidth]{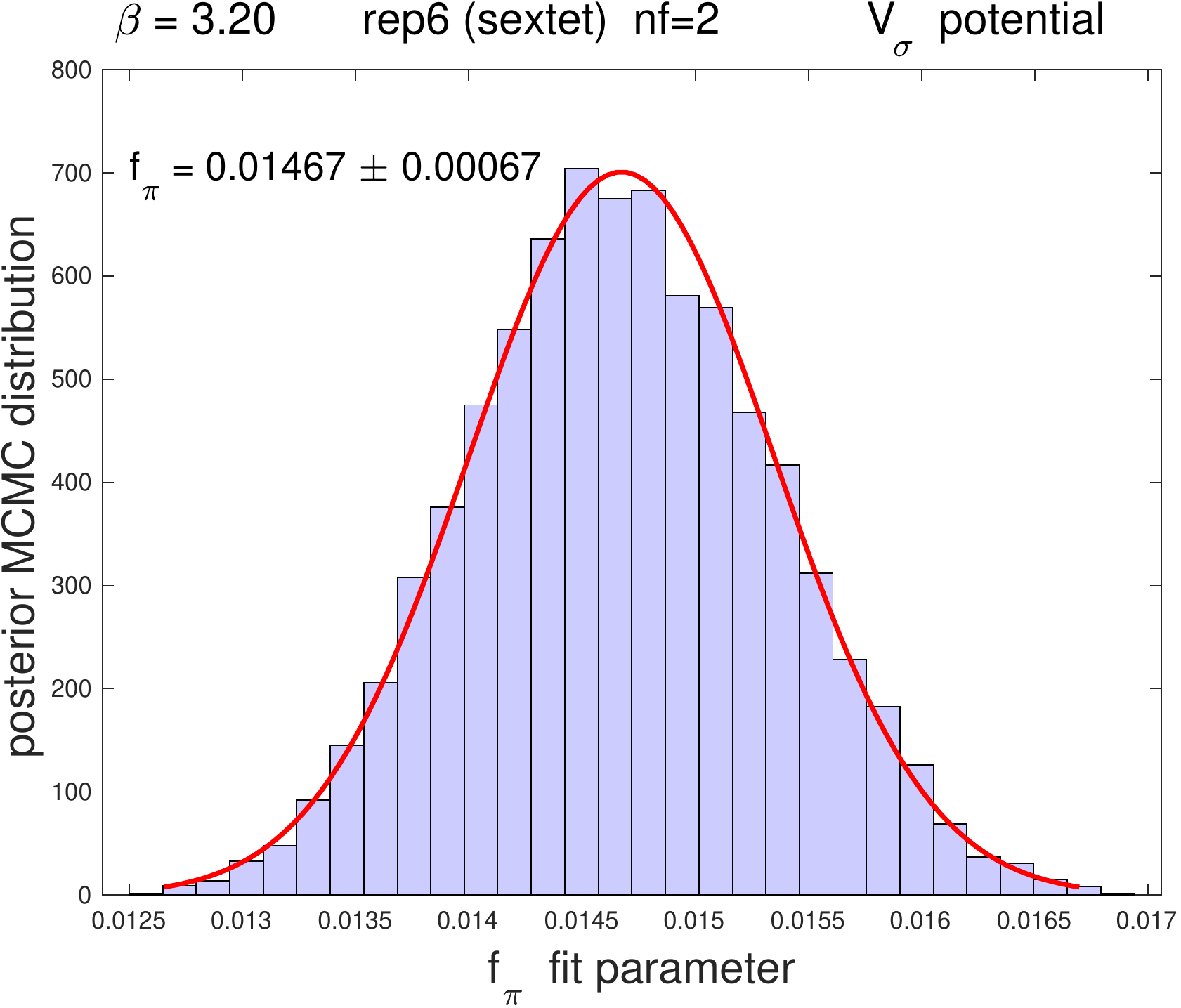}
		\end{subfigure}
		\begin{subfigure}{.49\linewidth}
			\caption*{\footnotesize posterior  $B_\pi$ for rep6  $V_\sigma$:}
			\includegraphics[width=\textwidth]{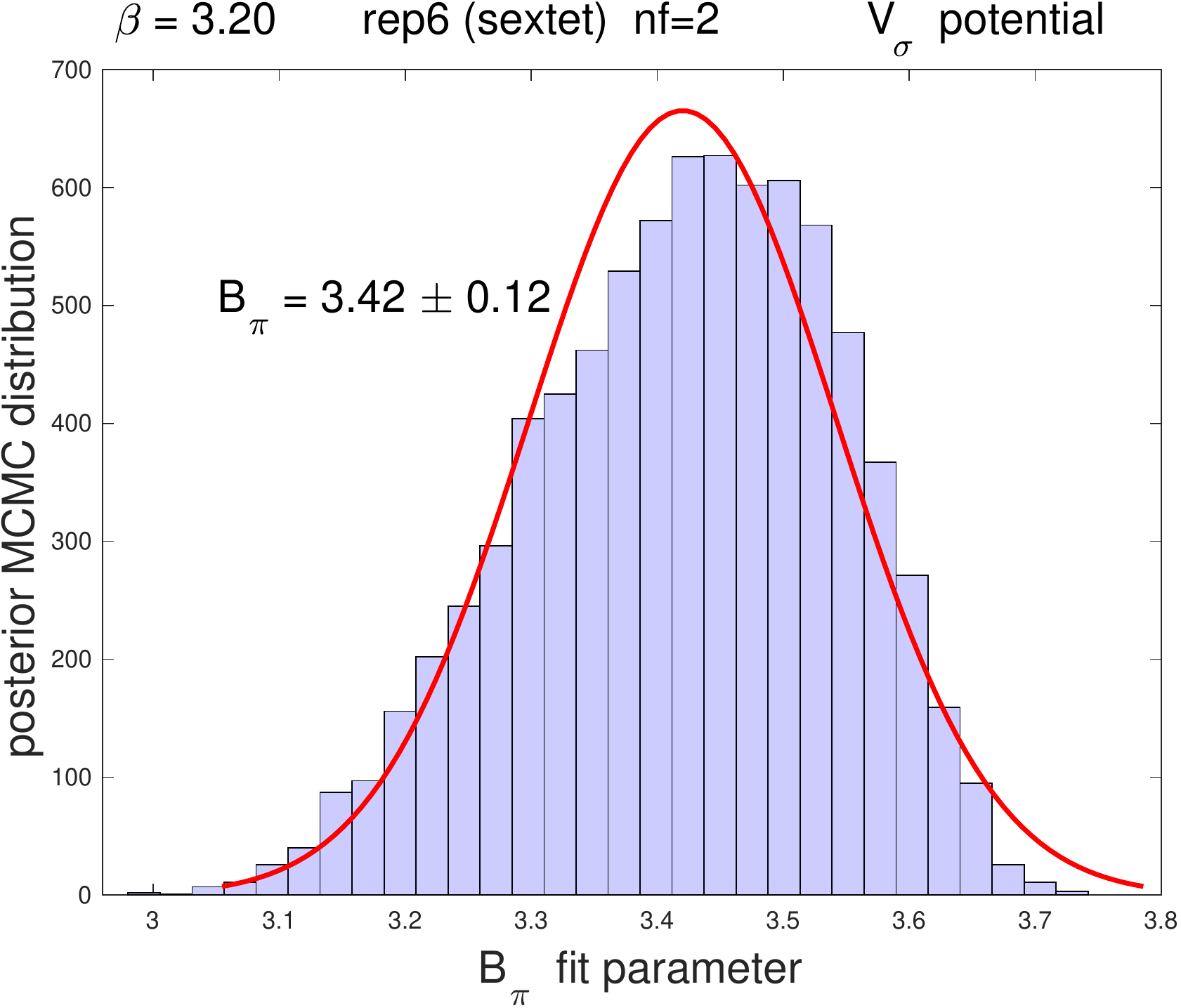}
		\end{subfigure}\\
		\begin{subfigure}{.49\linewidth}
			\caption*{\footnotesize posterior $\gamma$ for rep6  $V_\sigma$:}
			\includegraphics[width=\textwidth]{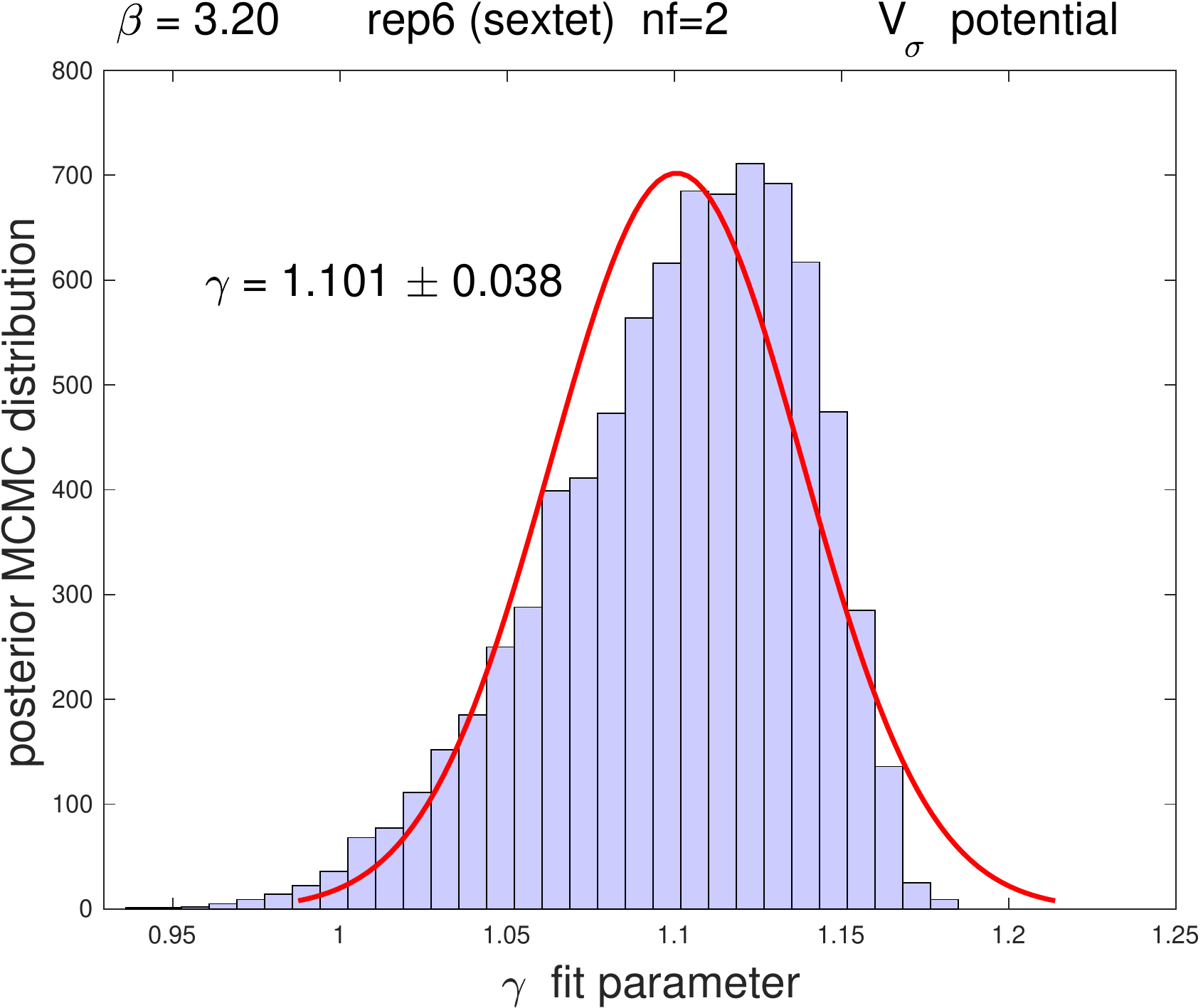}
		\end{subfigure}
		\begin{subfigure}{.49\linewidth}
			\caption*{\footnotesize posterior $m_d/f_\pi$ for rep6  $V_\sigma$:}
			\includegraphics[width=\textwidth]{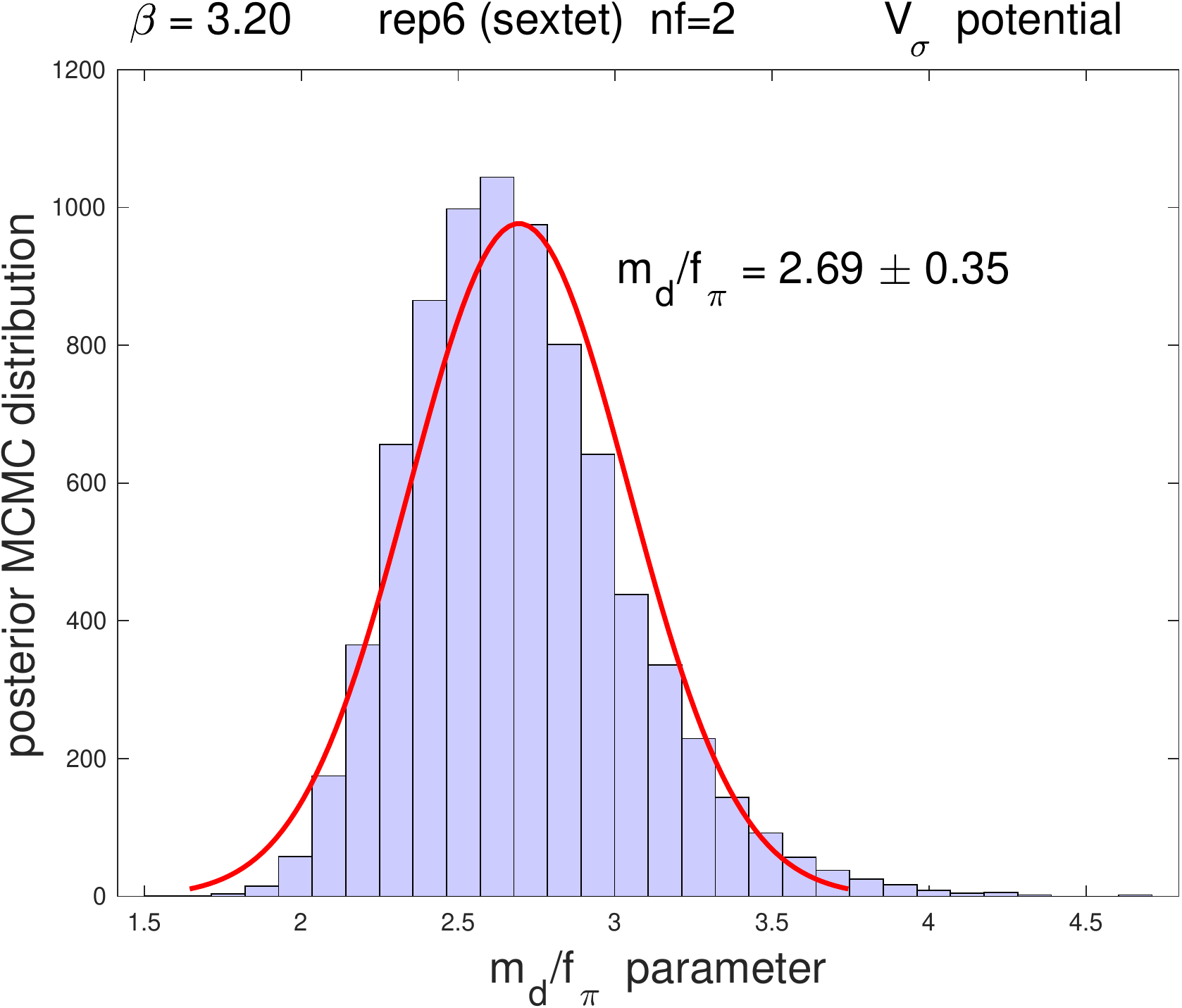}
		\end{subfigure}
	}
	\begin{subfigure}{.44\textwidth}
		\includegraphics[width=\textwidth]{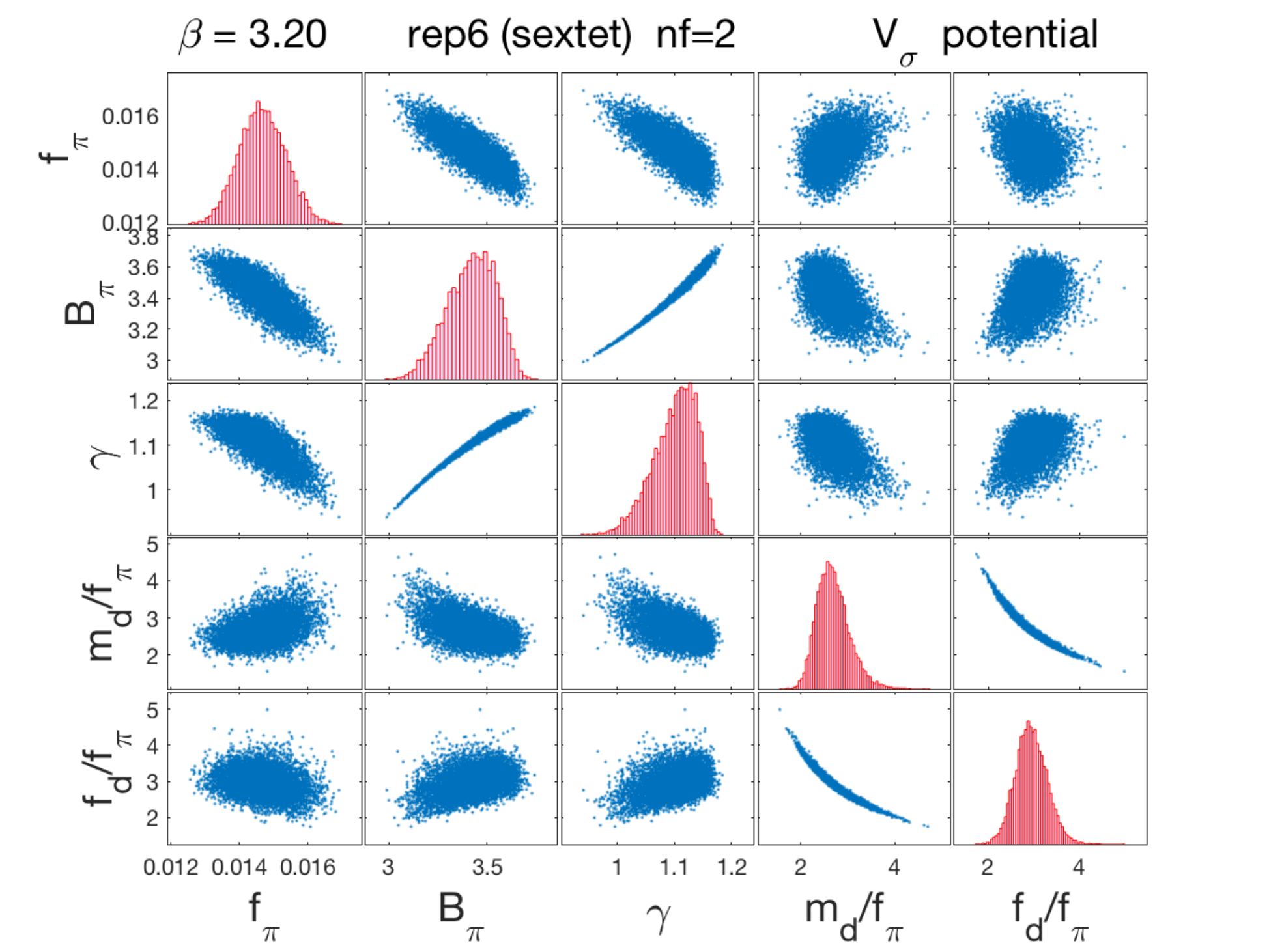}
		\caption*{\footnotesize Matrix plot of the five fitted physical parameters with their posterior histograms in the diagonal 
			and off-diagonal scatter plots of their correlations. Four of the histograms are also shown on the left with fitted means and
			1$\sigma$ equivalent percentile errors (the distributions are close to normal).}
	\end{subfigure}\hfill
	\vskip -0.1in
	\caption{\label{fig:rep6Vsigma}\footnotesize  MCMC based posterior probability distributions of the five fitted physical parameters 
		and their correlations for the $V_\sigma$ choice of the dilaton potential (rep6 sextet model). 
		Again, 	red lines indicate fits of normal distributions 
       to the histograms which show some deviations from the Gaussian shape as expected.
	}
\end{figure}
\noindent{\bf\em Physical parameters from dilaton EFT fits:}
The fitted posterior distributions of the physical parameters $f_\pi, B_\pi, y, m_d/f_\pi,f_d/f_\pi$ and their correlations 
are shown in Fig.~\ref{fig:rep6Vd} and Fig.~\ref{fig:rep6Vsigma}. Approximating the IML procedure, the distributions were generated in two stages.
At the first stage, $M_\pi(m)$, $F_\pi(m)$  correlated pairs of distribution functions were generated from extrapolation 
to infinite volume using finite size scaling (FSS) analysis separately at each fermion mass.
Lattice input to the FSS fits was provided by three pairs of six $M_\pi(m,L,L_t)$, $F_\pi(m,L,L_t)$ data
from three volumes with linear sizes $L,L_t$ in respective spatial and time directions. Maximum Likelihood FSS fits 
gave us correlated infinite volume $M_\pi(m)$, $F_\pi(m)$ pairs with covariance matrices which were used 
in Markov Chain Monte Carlo (MCMC) samples to generate the posterior probability distributions of the $M_\pi(m)$, $F_\pi(m)$ pairs as input
for the second stage of the fitting procedure.  $M_d(m)$ probability distributions for input to the second stage were generated without FSS from 
normal distributions with previously determined variances at the largest available volume for each $m$.
For each drawing from the $M_\pi(m)$, $F_\pi(m), M_d(m)$ distributions, 12 input data were selected using $m=0.015/0.020/0.030/0.040$ 
for the non-linear fitting procedure of the five physical parameters from 
respective Eqs.(\ref{eq:Cd}-\ref{eq:Md1}) of the two dilaton potentials.
We checked that adding the lowest fermion mass $m=0.0010$ with incomplete FSS for $F_\pi$ would not have any 
significant effect on the $M_\pi(m)$, $F_\pi(m)$ distributions or the posterior distributions of the fitted physical parameters at the second stage.
The posterior distributions of the five fitted physical parameters and their correlations were generated from the order of ten thousand drawings from the 
$M_\pi(m), F_\pi(m), M_d(m)$ distributions as shown in Fig.~\ref{fig:rep6Vd} and Fig.~\ref{fig:rep6Vsigma}. 

Results for the $V_d(\chi)$ choice of the dilaton potential in Eq.(\ref{eq:Vd}) have some remarkable features with important implications 
for added post-conference analysis.
The dilaton mass $m_d/f_\pi = 1.56(28)$ is dramatically lower than  extrapolated results from chiral perturbation theory with less control. If confirmed, this light dilaton state alone could change our perspective on the sextet model for future investigations.  The anomalous dimension $\gamma = 0.870(93)$ is consistent with direct determination from the renormalized mode number distribution of the Dirac operator. The estimate of $f_\pi = 0.0109(21)$ is lower than what was obtained in earlier $\chi PT$ fits which might require the recalibration of the  separation between the $0^{++}$ scalar and the associated heavy resonance spectrum. 
The value of $B_\pi = 2.76(28)$ is close to what was determined from $\chi PT$ as shown in Fig.~\ref{fig:rep6chiPT}. However, the result  $f_d/f_\pi = 2.94(35)$ would present phenomenological difficulties for potential BSM  applications. 

Perhaps not surprisingly, results from the choice $V_\sigma(\chi)$  in Eq.(\ref{eq:Vsigma}) are closer to what was expected from $\sigma$-model inspired ${\chi PT}$ estimates. The value of $B_\pi = 3.42(12)$ is practically the same as what we obtained from chiral log fits in Section~\ref{section:su2} and the dilaton mass  is heavier in $f_\pi$ units, $m_d/f_\pi = 2.69(35)$, closer to what we expected earlier for the $0^{++}$ scalar.  The ratio $f_d/f_\pi = 3.22(37)$ is close to what we obtained with the choice of $V_d(\chi)$, with similar challenges for Electroweak embedding. 

Sensitivity of the light scalar mass to the choice of the dilaton potential is the most striking outcome of the dilaton analysis in the sextet model. 
This might require revised strategies in future work including questions on the size of corrections to the tree-level approximation, affected by 
the small value of $f_\pi$. It is interesting to note that the $V_d(\chi)$ choice from Eq.(\ref{eq:Vd}) is  preferred near the CW from 
theoretical arguments in~\cite{Golterman:2016lsd}, not directly applicable to the sextet model. 
\subsection{Dilaton EFT fits from ${\mathbf V_\sigma}$ and ${\mathbf V_d}$  dilaton potentials:  $\mathbf n_f=8$ model}
\begin{figure}[h!]
	\centering
	\parbox{.53\textwidth}{
		\begin{subfigure}{.49\linewidth}
			\caption*{\footnotesize $n_f=8~~~ f_\pi$ fit:}
			\includegraphics[width=\textwidth]{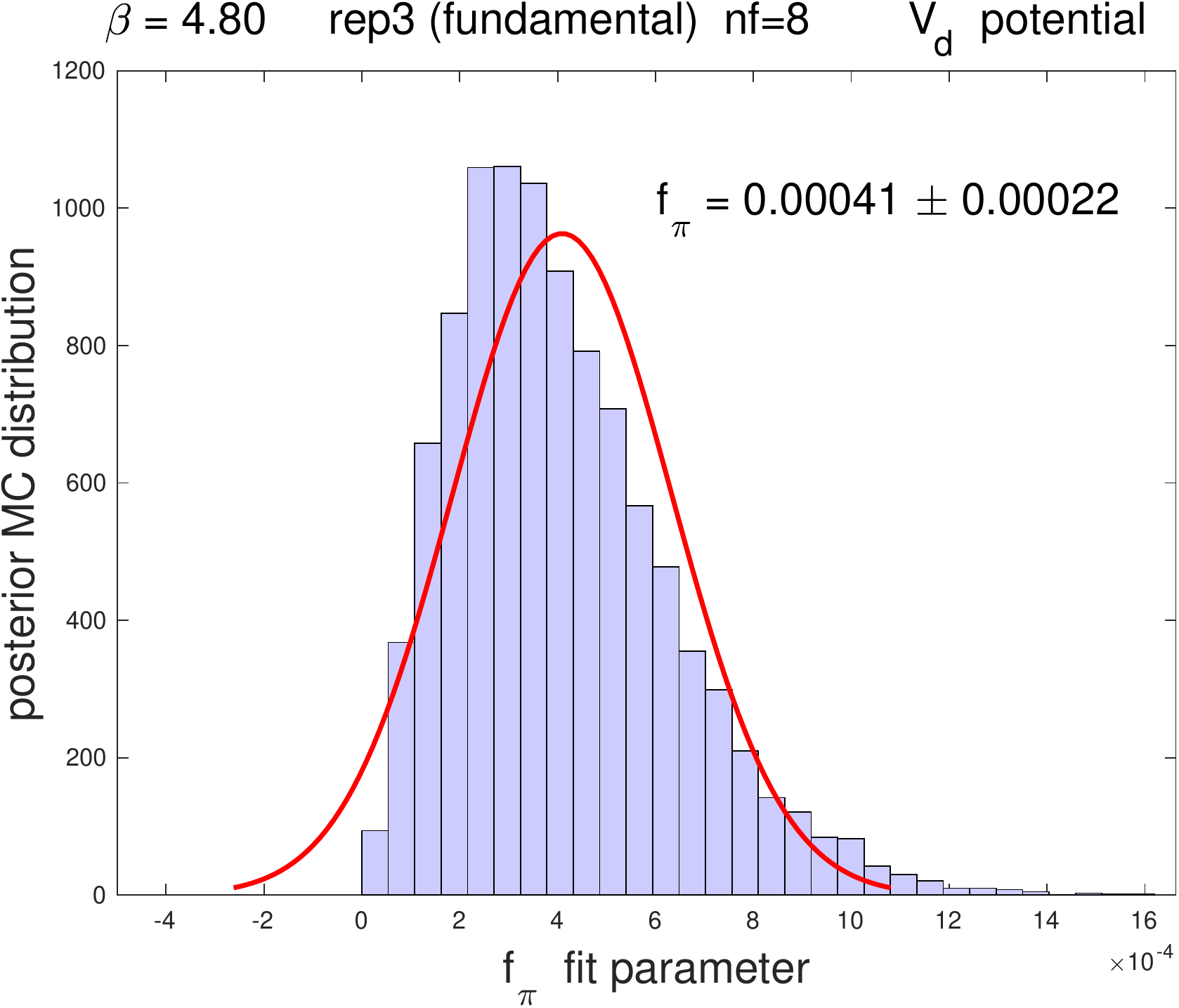}
		\end{subfigure}
		\begin{subfigure}{.49\linewidth}
			\caption*{\footnotesize $n_f=8~~~ B_\pi$ fit:}
			\includegraphics[width=\textwidth]{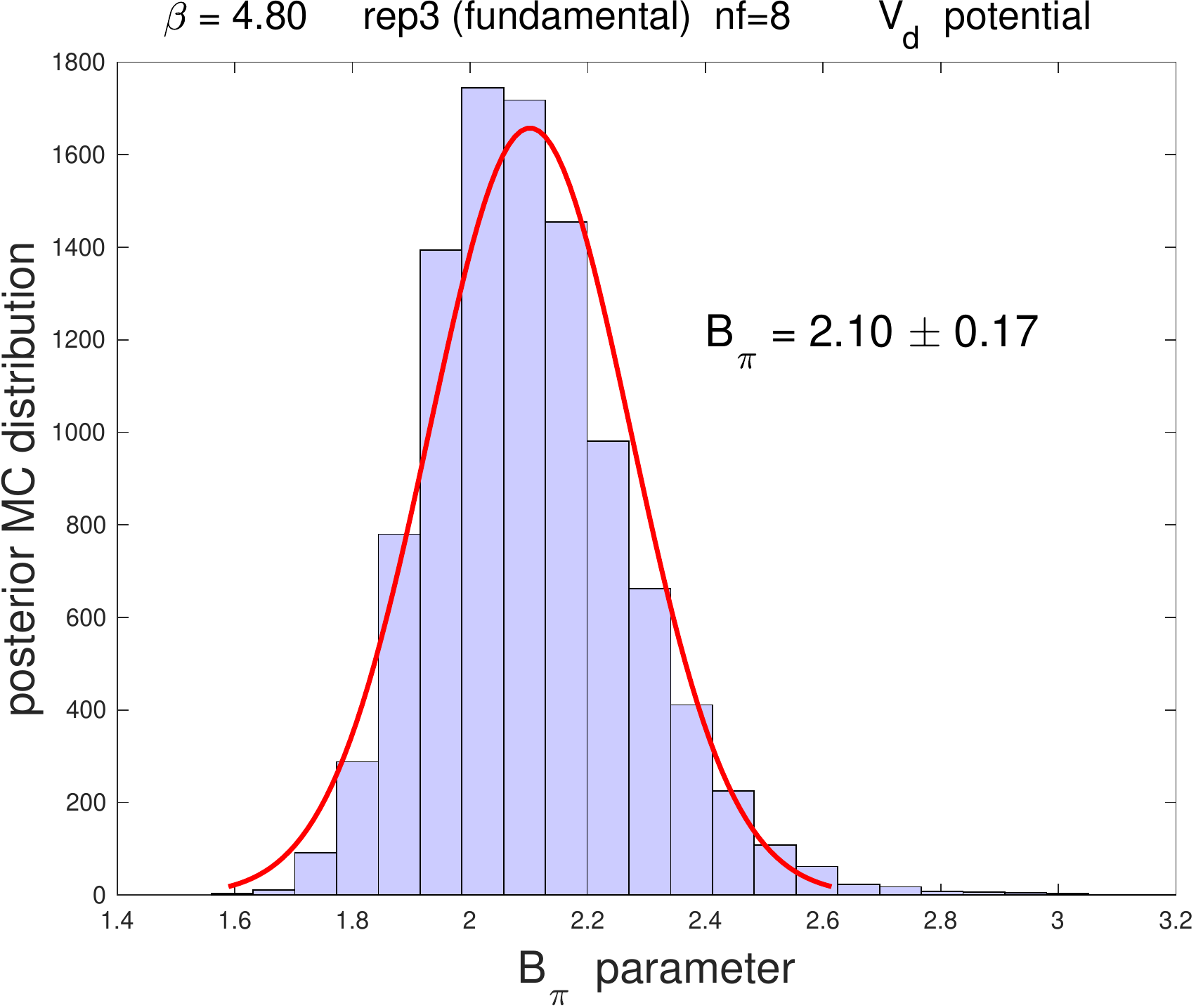}
		\end{subfigure}\\
		\begin{subfigure}{.49\linewidth}
			\caption*{\footnotesize $n_f=8~~~\gamma$ fit:}
			\includegraphics[width=\textwidth]{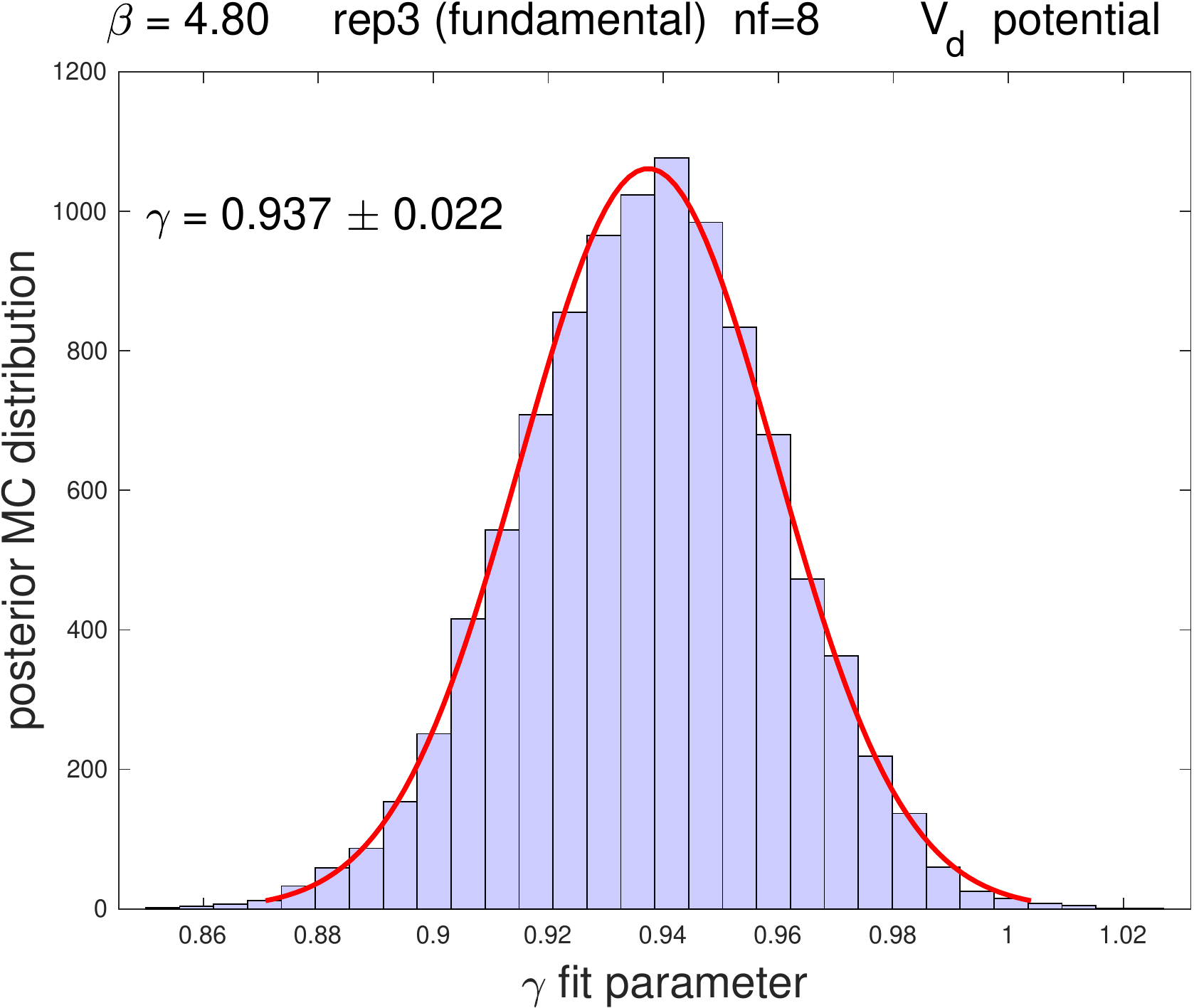}
		\end{subfigure}
		\begin{subfigure}{.49\linewidth}
			\caption*{\footnotesize $n_f=8~~~ m_d/f_\pi\cdot f_d/f_\pi~$ fit:}
			\includegraphics[width=\textwidth]{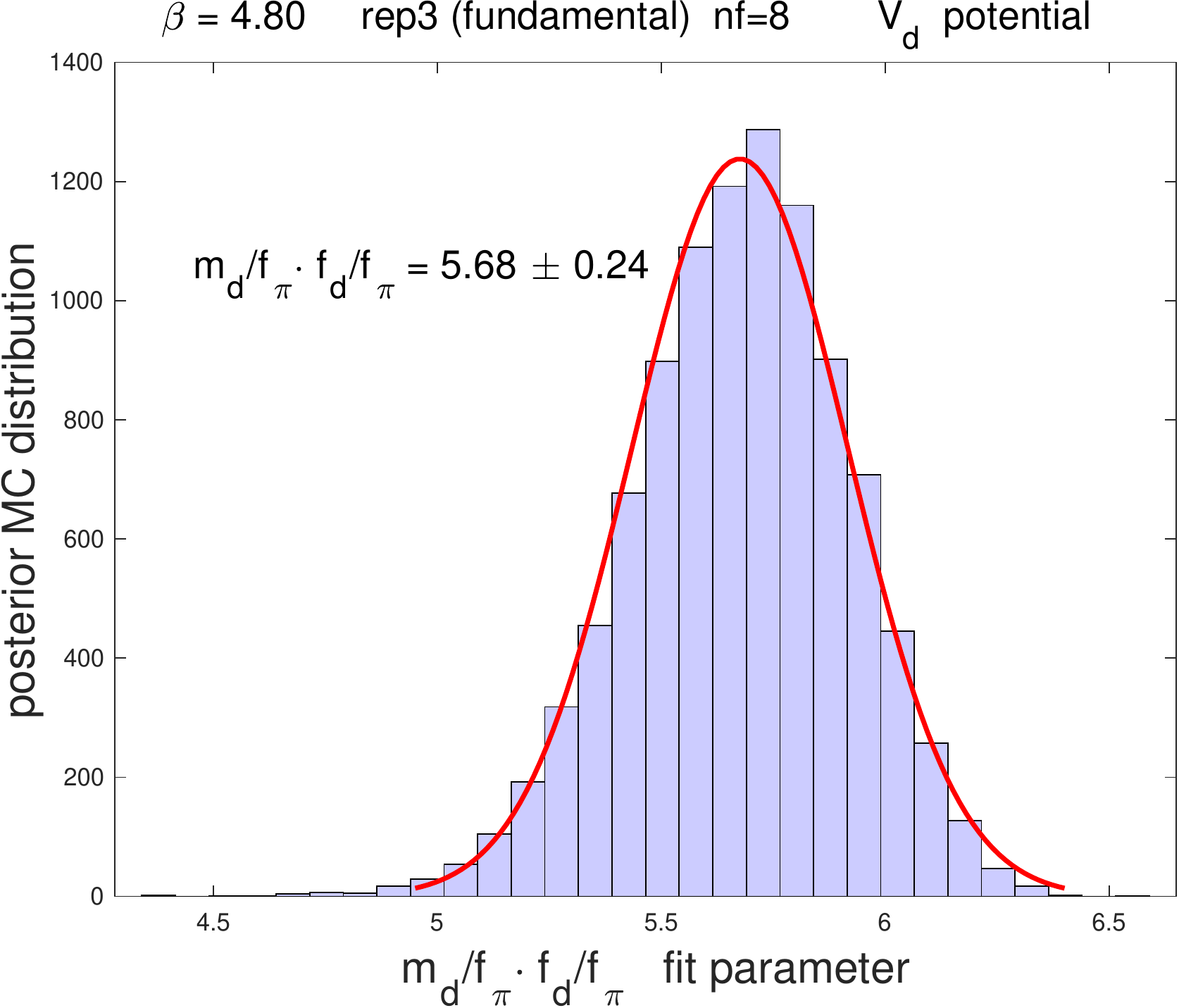}
		\end{subfigure}
	}
	\begin{subfigure}{.44\textwidth}
		\includegraphics[width=\textwidth]{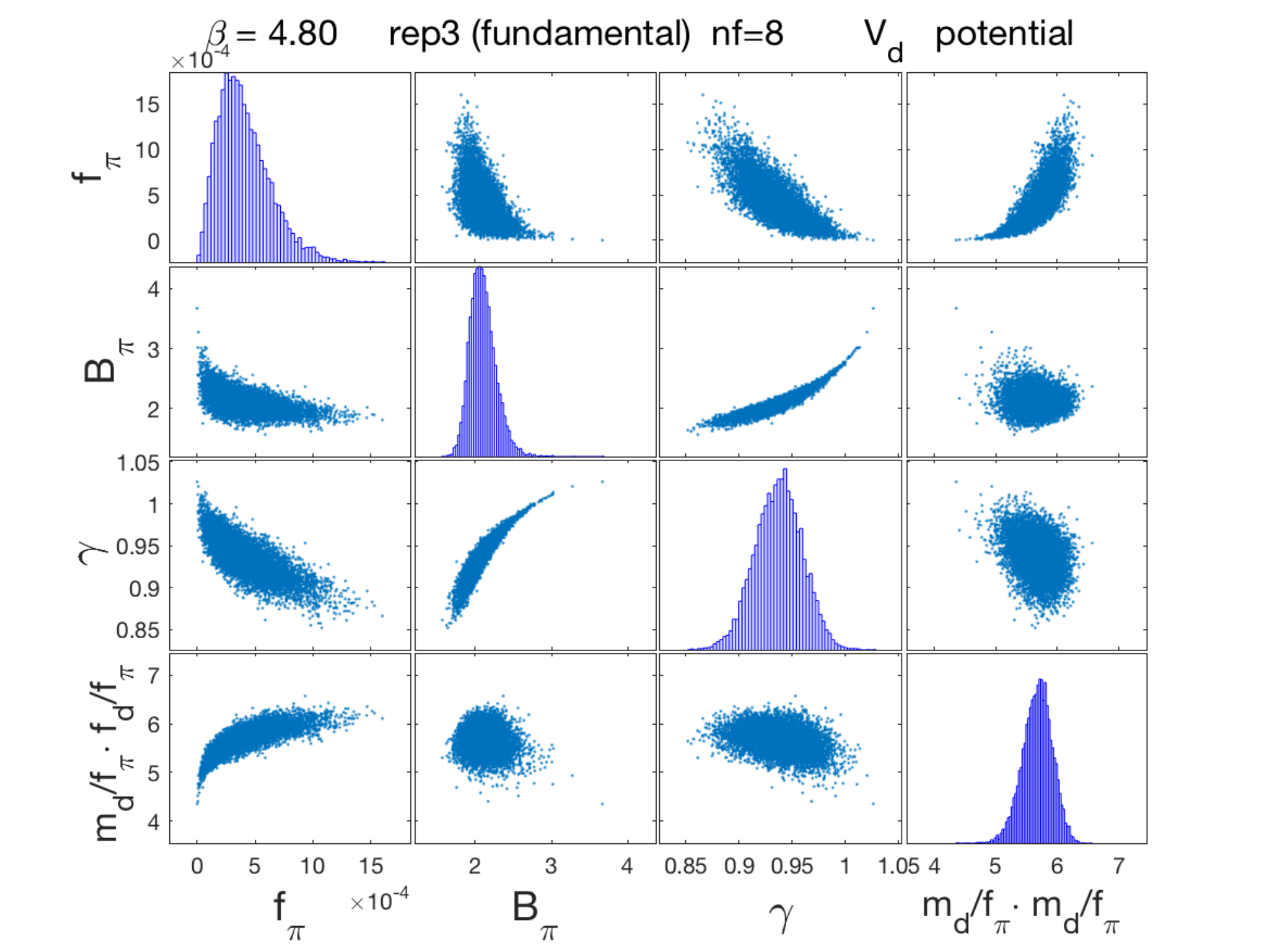}
		\caption*{\footnotesize Matrix plot of the fitted physical parameters with histogram in the diagonal 
			and off-diagonal scatter plots of their correlations. The histograms are also shown on the left with fitted means and
			1$\sigma$ equivalent percentile errors (the $f_\pi$ distribution deviates from normal). }
	\end{subfigure}\hfill
	\caption{\footnotesize  Histograms of the four fitted physical parameters 
		and scatter plots of their correlations for the $V_d$ choice of the dilaton potential (rep3 $n_f=8$ model). }\label{fig:rep3Vd}
	\vskip -0.1in
\end{figure}

\begin{figure}[h]
	\centering
	\parbox{.53\textwidth}{
		\begin{subfigure}{.49\linewidth}
			\caption*{\footnotesize  $n_f=8~~~f_\pi$ fit:}
			\includegraphics[width=\textwidth]{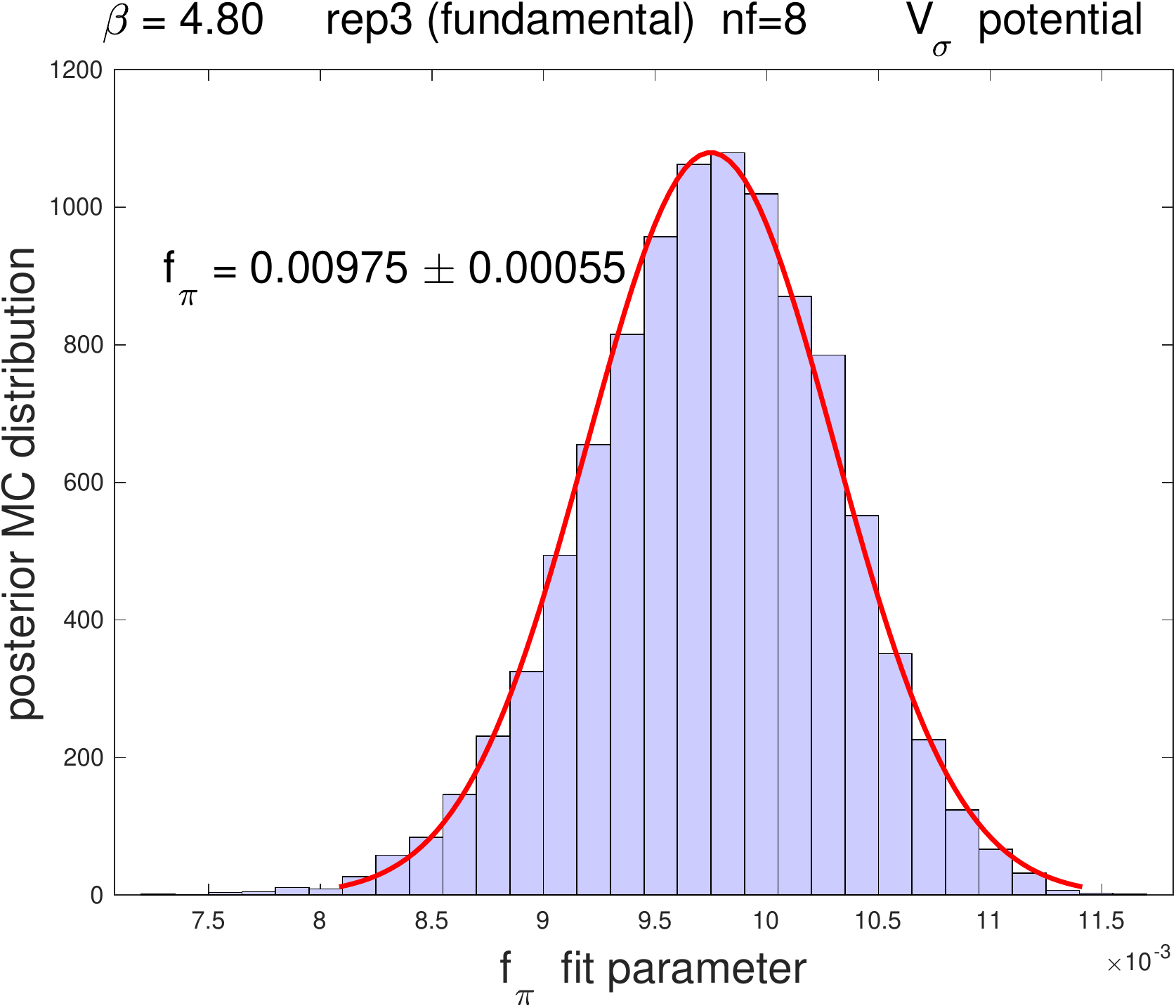}
		\end{subfigure}
		\begin{subfigure}{.49\linewidth}
			\caption*{\footnotesize  $n_f=8~~~B_\pi$ fit:}
			\includegraphics[width=\textwidth]{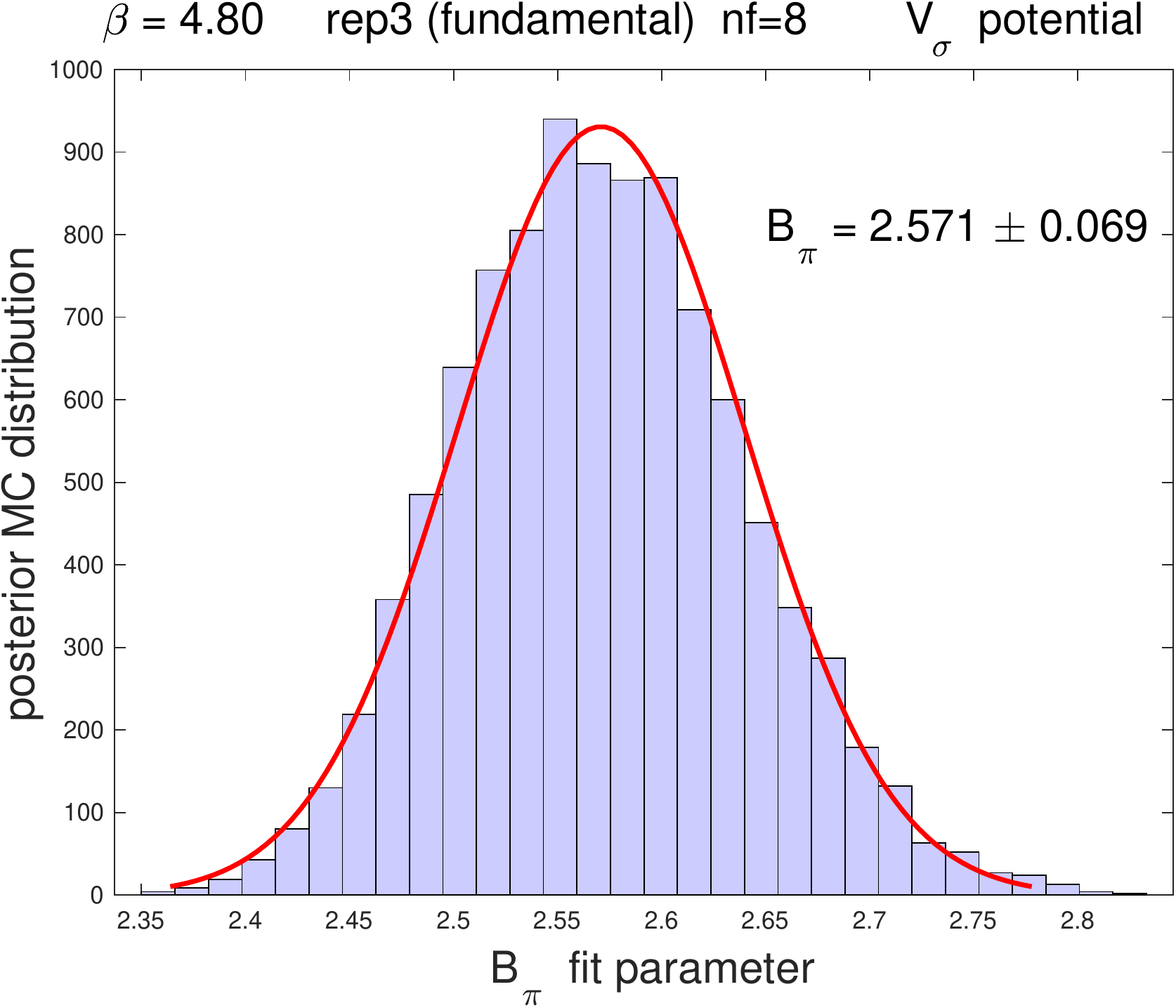}
		\end{subfigure}\\
		\begin{subfigure}{.49\linewidth}
			\caption*{\footnotesize $n_f=8~~~\gamma$ fit:}
			\includegraphics[width=\textwidth]{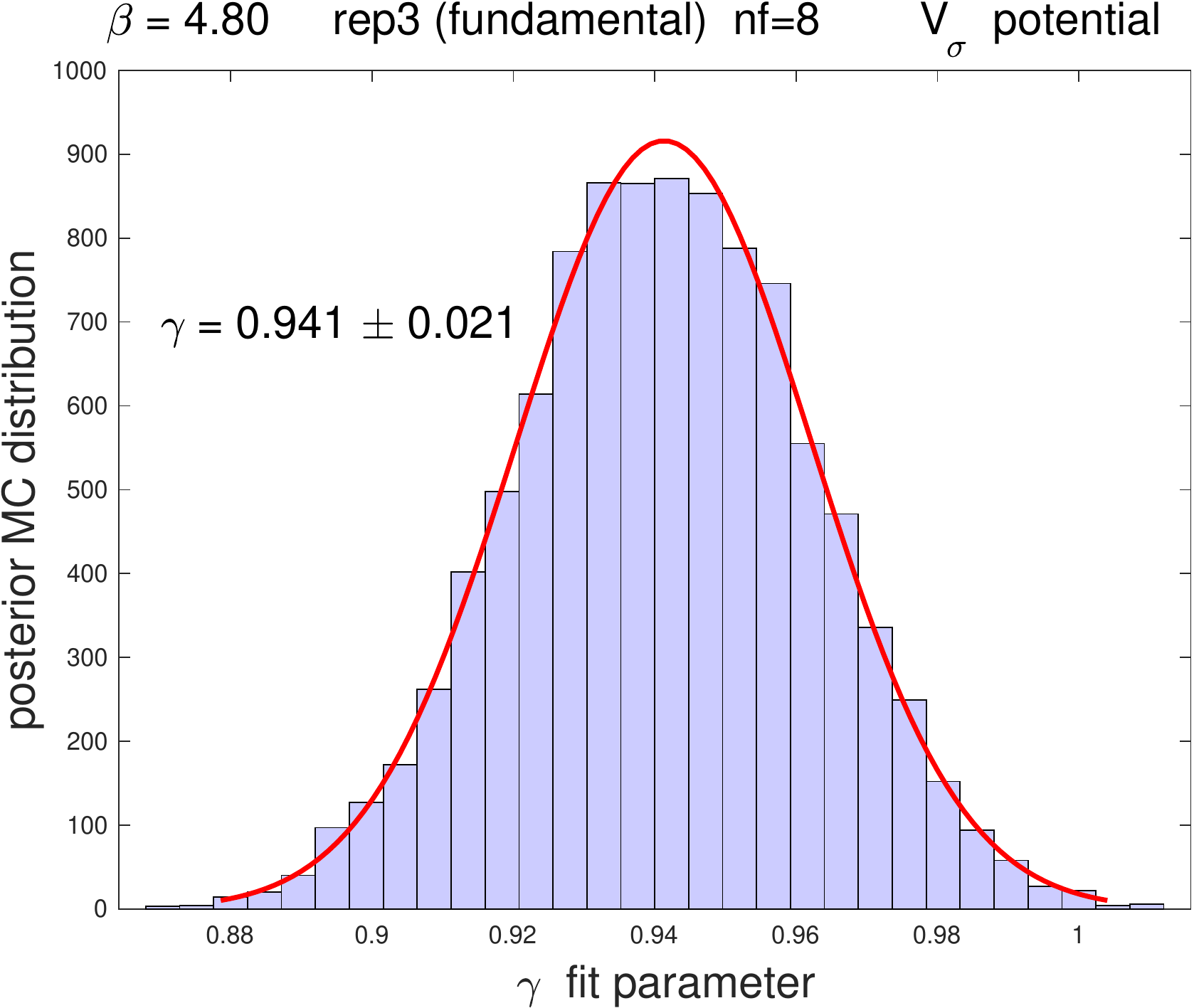}
		\end{subfigure}
		\begin{subfigure}{.49\linewidth}
			\caption*{\footnotesize  $n_f=8~~~m_d/f_\pi\cdot f_d/f_\pi~$ fit:}
			\includegraphics[width=\textwidth]{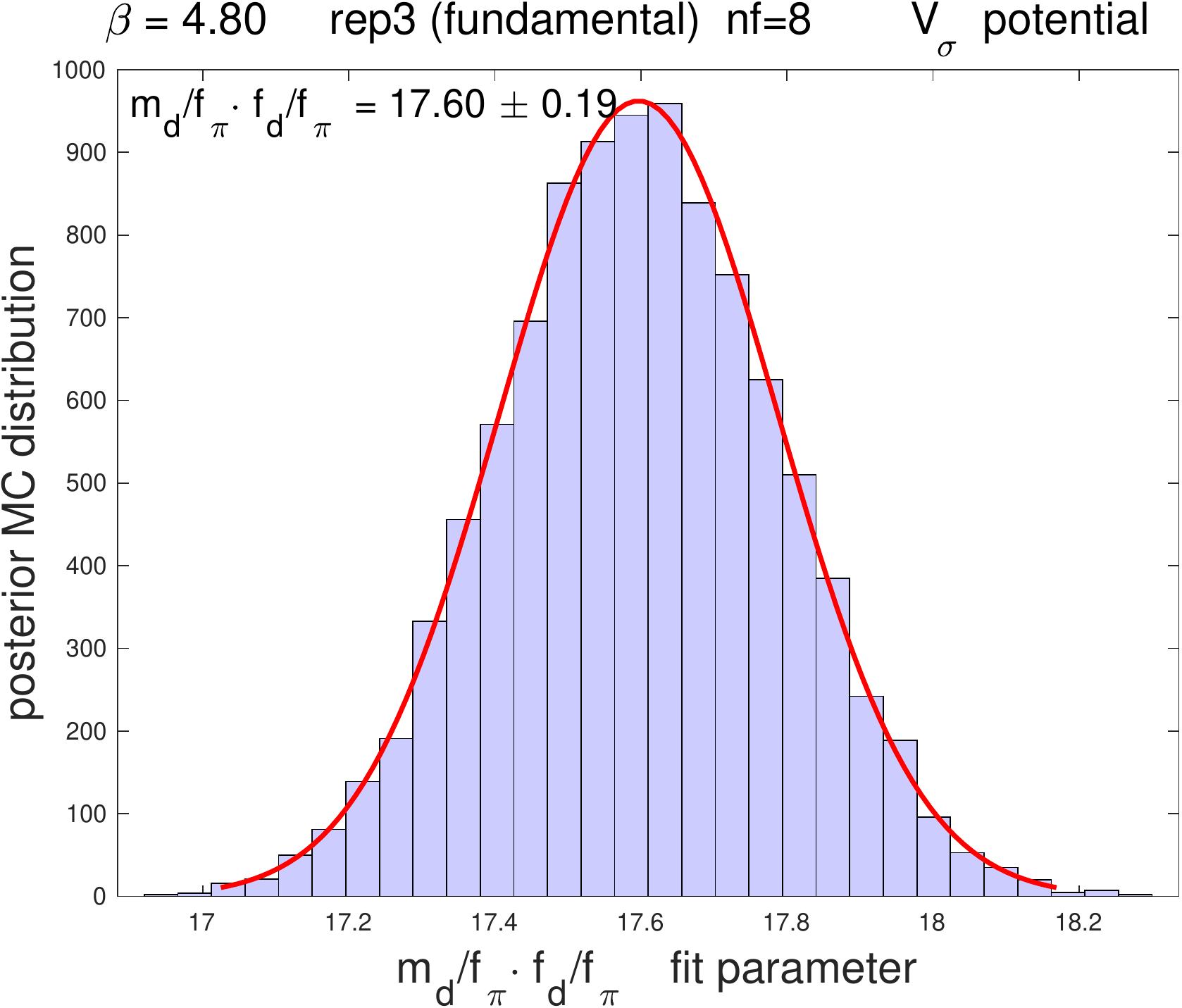}
		\end{subfigure}
	}
	\begin{subfigure}{.44\textwidth}
	\includegraphics[width=\textwidth]{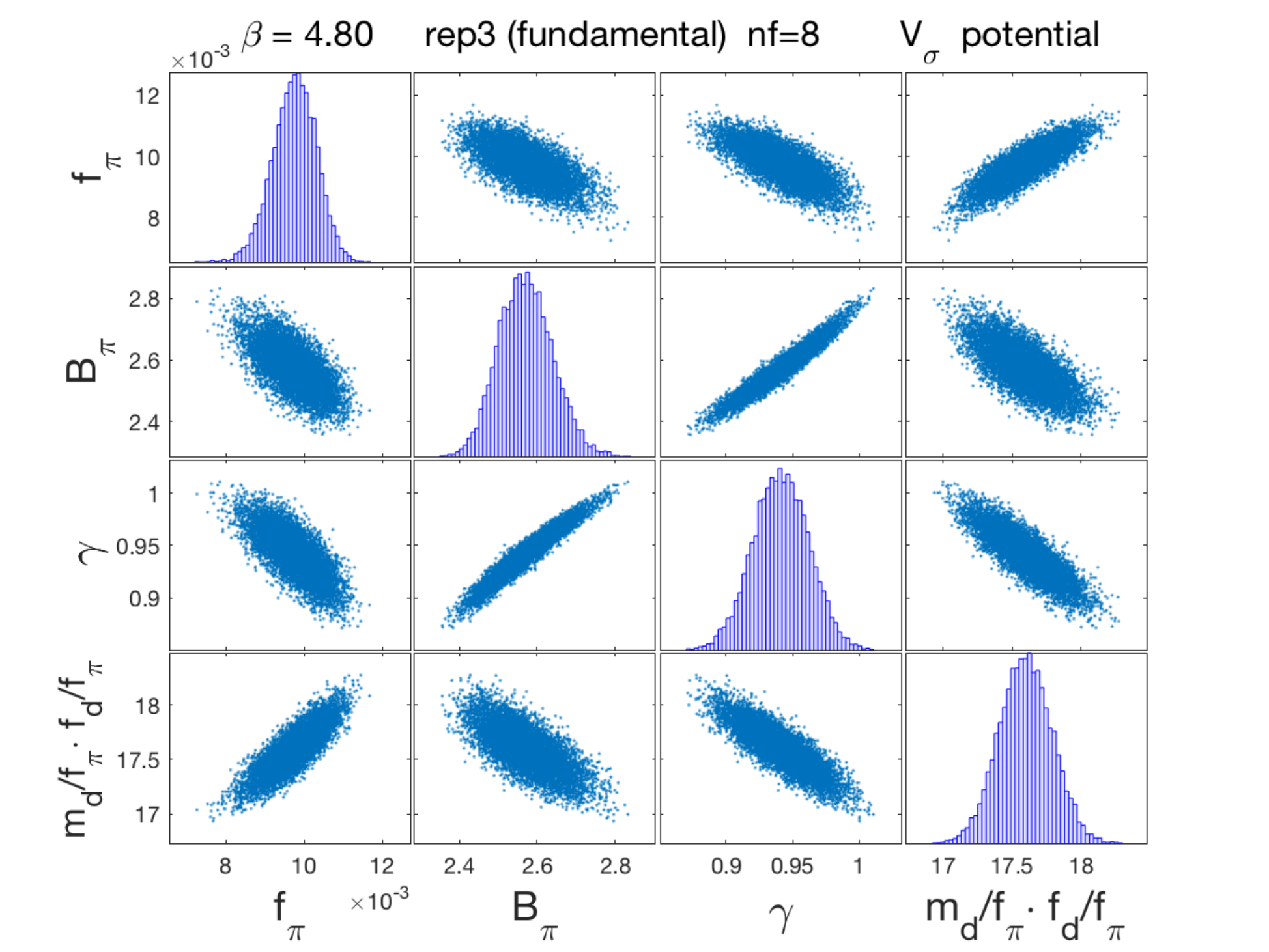}
	\caption*{\footnotesize Matrix plot of the four fitted physical parameters with histograms in the diagonal 
		and off-diagonal scatter plots of their correlations. The histograms are also shown on the left with fitted means and
		1$\sigma$ equivalent percentile errors (the fitted distributions are close to normal). } 
\end{subfigure}\hfill
	\caption{\footnotesize Histograms of four fitted physical parameters 
		and scatter plots of their correlations for the $V_\sigma$ choice of the dilaton potential (rep3 $n_f=8$ model). }\label{fig:rep3Vsigma}
\end{figure}

With input data from~\cite{Appelquist:2018yqe}, the histograms of four fitted physical parameters $f_\pi, B_\pi, y, m_d/f_\pi\cdot f_d/f_\pi$ and their correlations 
are shown in Fig.~\ref{fig:rep3Vsigma} and Fig.~\ref{fig:rep3Vd}. Using the largest volumes from~\cite{Appelquist:2018yqe} 
did not allow us for MCMC based FSS, instead normal distributions of $M_\pi,F_\pi$ were inputs to the fitting
procedure.
After experimenting with  five-parameter fits, we
reduced the analysis to four physical parameters without $M_d$ input.
Accordingly, we did not use Eq.(\ref{eq:Md}) and  (\ref{eq:Md1}) with inputs from  $M_d$ in~\cite{Appelquist:2018yqe}, 
and only the product $m_d/f_\pi\cdot f_d/f_\pi$ was fitted without separating  $m_d/f_\pi$.

The $V_\sigma$ based $n_f=8$ fits do not show unexpected features. On the other hand, we had difficulties to interpret 
the $V_d$ based dilaton fits, in particular the very low value of $f_\pi = 0.00041(22)$ calling into question the leading tree-level 
approximation to $V_d$ based  tests of the dilaton EFT. We will return to investigate some stable 
way of including the scalar mass $M_d$ directly in the fitting procedure.  

There is, however, an important aspect of dilaton EFT based fits to the $n_f=8$ model which is not affected by details of our  fitting procedure.  
It was argued in~\cite{Golterman:2018mfm} and presented at the conference~\cite{Golterman:2018bpc} that the analysis 
of the $n_f=8$ model in~\cite{Appelquist:2017wcg,Appelquist:2017vyy} is based on input data in the high-mass range of fermion mass deformations
where $M_\pi$ and $F_\pi$ would show conformal scaling. It was estimated in~\cite{Golterman:2018mfm,Golterman:2018bpc} 
that two orders of magnitude drop would be required from the currently available  fermion mass range before the 
onset of chiral behavior is reached at very low fermion masses, outside the reach of realistic lattice simulation. 
The argument was based on Eq.(\ref{eq:EFT}) of the EFT for the choice $V_d$ of the dilaton potential in Eq.(\ref{eq:Vd}). 
\begin{figure}[htb!]
	\begin{center}
		\begin{tabular}{ccc}
			\includegraphics[width=0.3\textwidth]{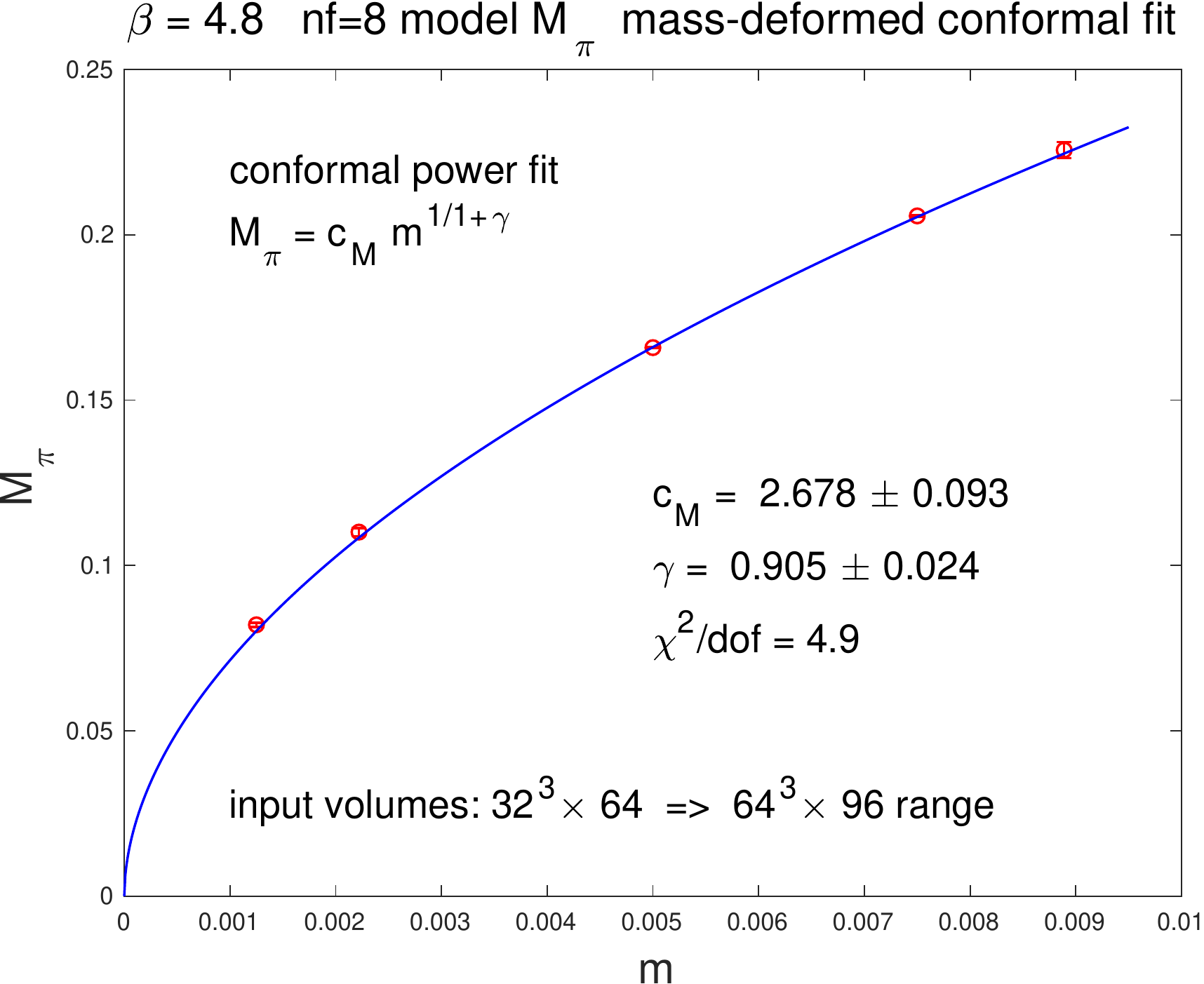}&
            \includegraphics[width=0.3\textwidth]{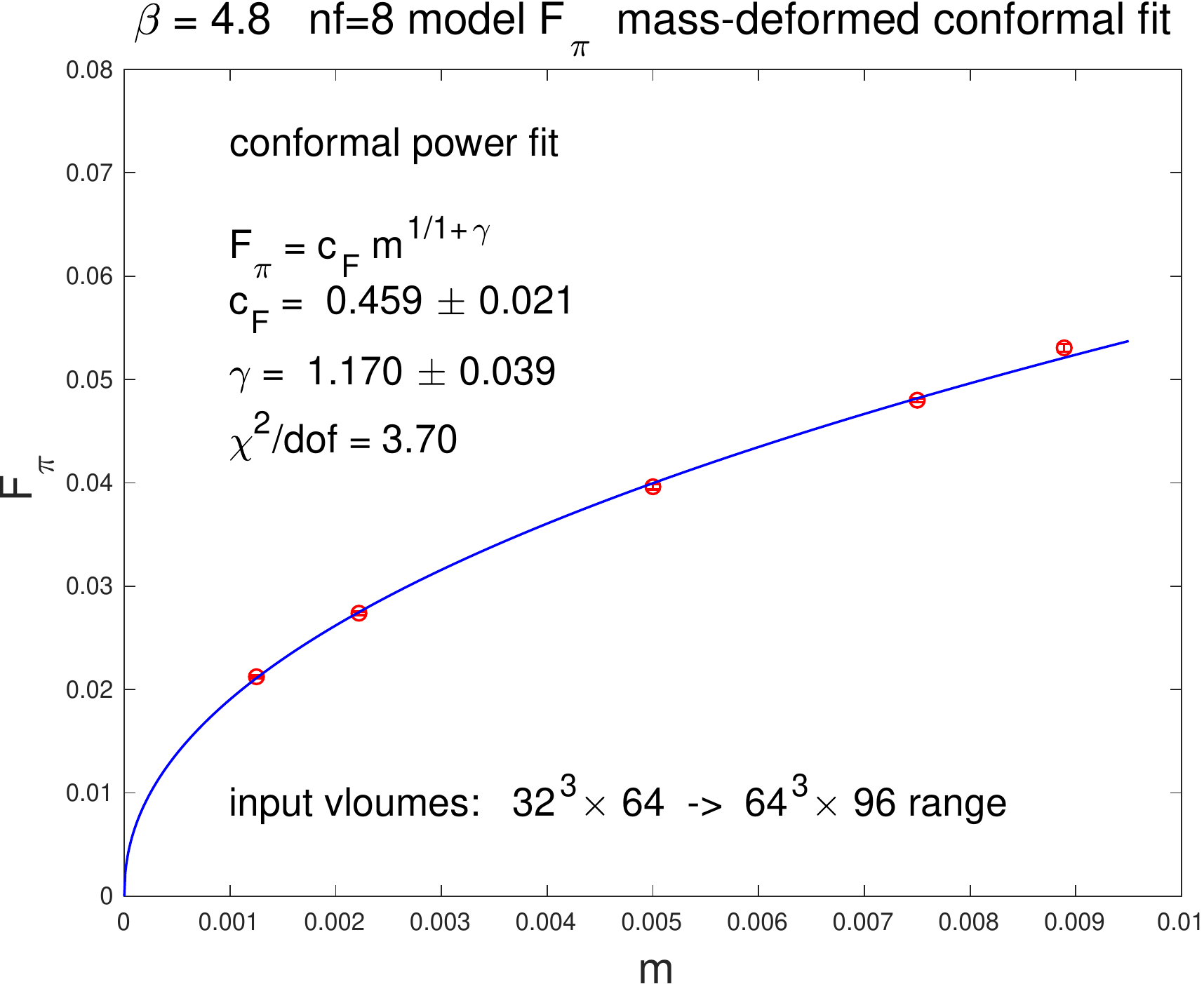}&
            \includegraphics[width=0.3\textwidth]{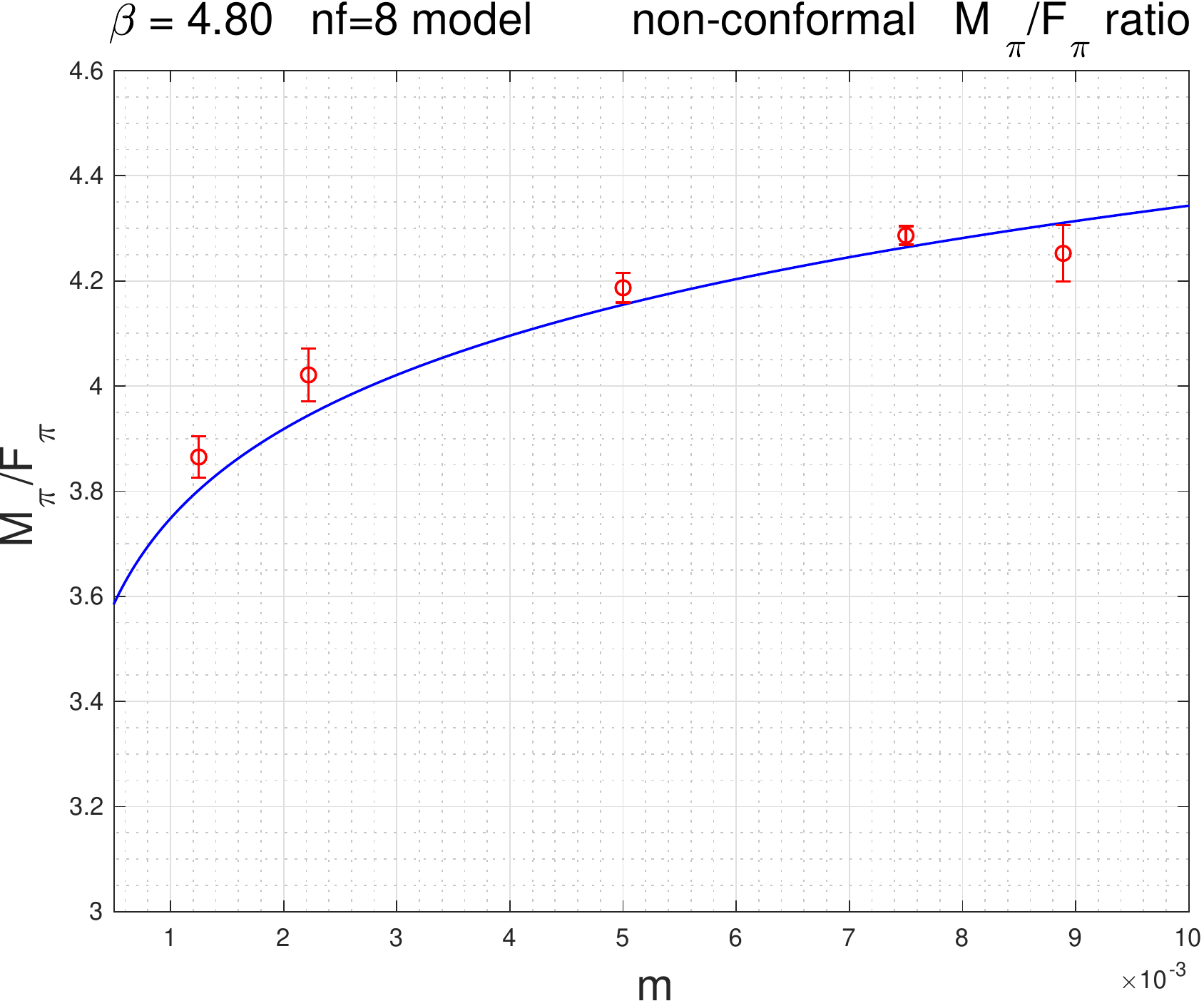}		
		\end{tabular}
	\end{center}
	\vskip -0.25in
	\caption{\footnotesize Inconsistencies of the conformal tests in the $n_f=8$ model, exhibiting considerable variation of the  $M_\pi/F_\pi$ ratio 
		in the fitted fermion mass range from incompatibility conformal exponents between $M_\pi$ and $F_\pi$.}
	\label{fig:rep3conform}
\end{figure}
We checked our input data set $M_\pi,F_\pi$ from the recent LSD publication~\cite{Appelquist:2018yqe} for conformal behavior 
in the fermion mass range of the input data. The results in Fig.~\ref{fig:rep3conform}  for $M_\pi$ and $F_\pi$ do not show conformal behavior in 
the mass range of the dilaton analysis with considerable variation of the  $M_\pi/F_\pi$ ratio and inconsistent 
conformal $\gamma$ exponents from forced conformal fits of $M_\pi$ and $F_\pi$. 
Although the identification of current $n_f=8$ simulations with the high-mass conformal regime will require correction terms~\cite{Golterman:2018bpc},
the estimated  two order of magnitude drop in $m$ for reaching the regime of $\chi SB$  might not be far from what is required.
At the conference we presented an idea how this large drop in $m$ might be reached instead in the $\epsilon$-regime of $\chi SB$.
\vskip -0.5in
\section{Dilaton EFT analysis in the $\mathbf{\epsilon}$-regime and RMT}\label{section:rmt}

\begin{figure}[htb!]
	\begin{center}
		\begin{tabular}{cc}
			\includegraphics[width=0.43\textwidth]{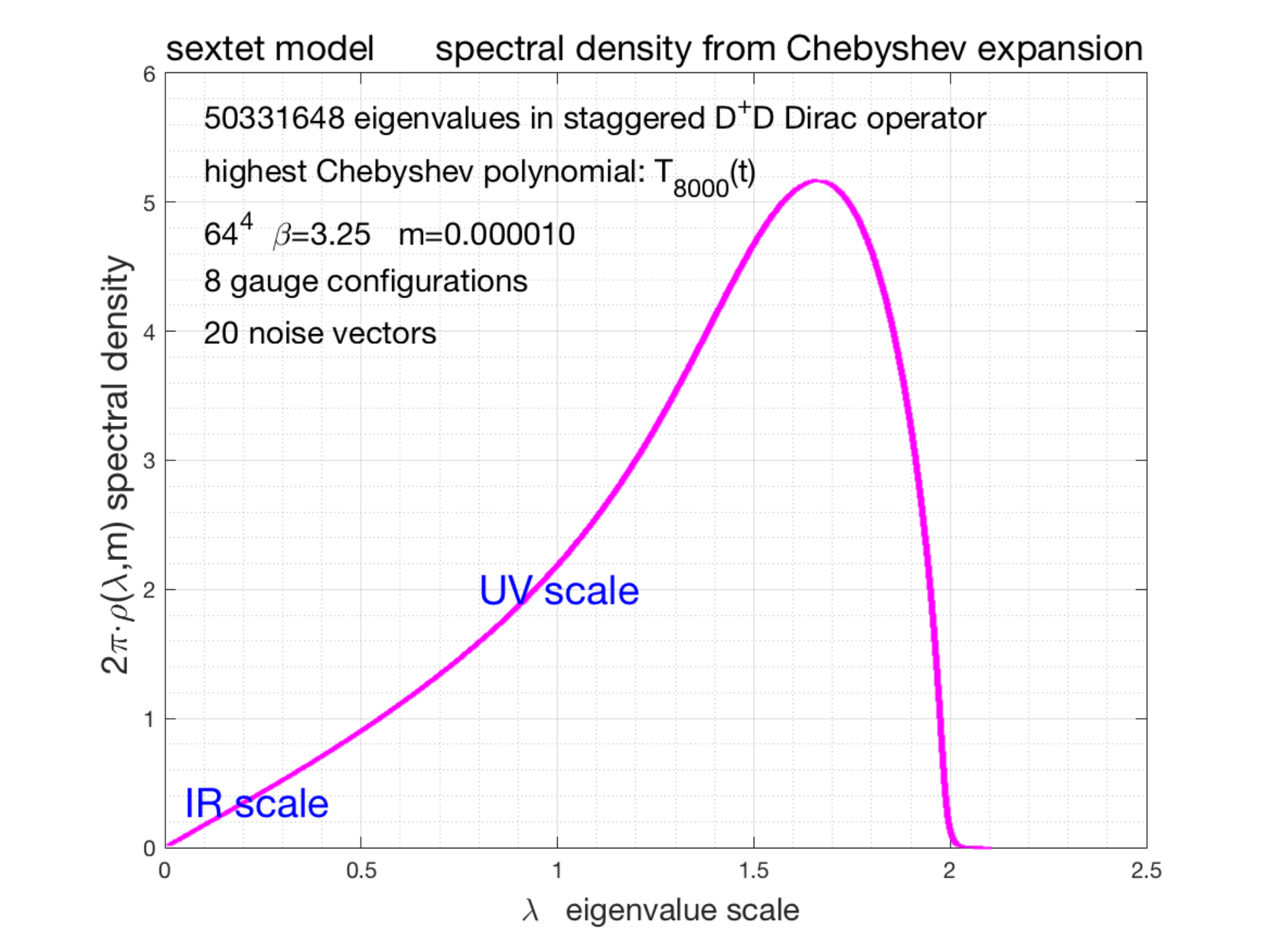}&
			\includegraphics[width=0.37\textwidth]{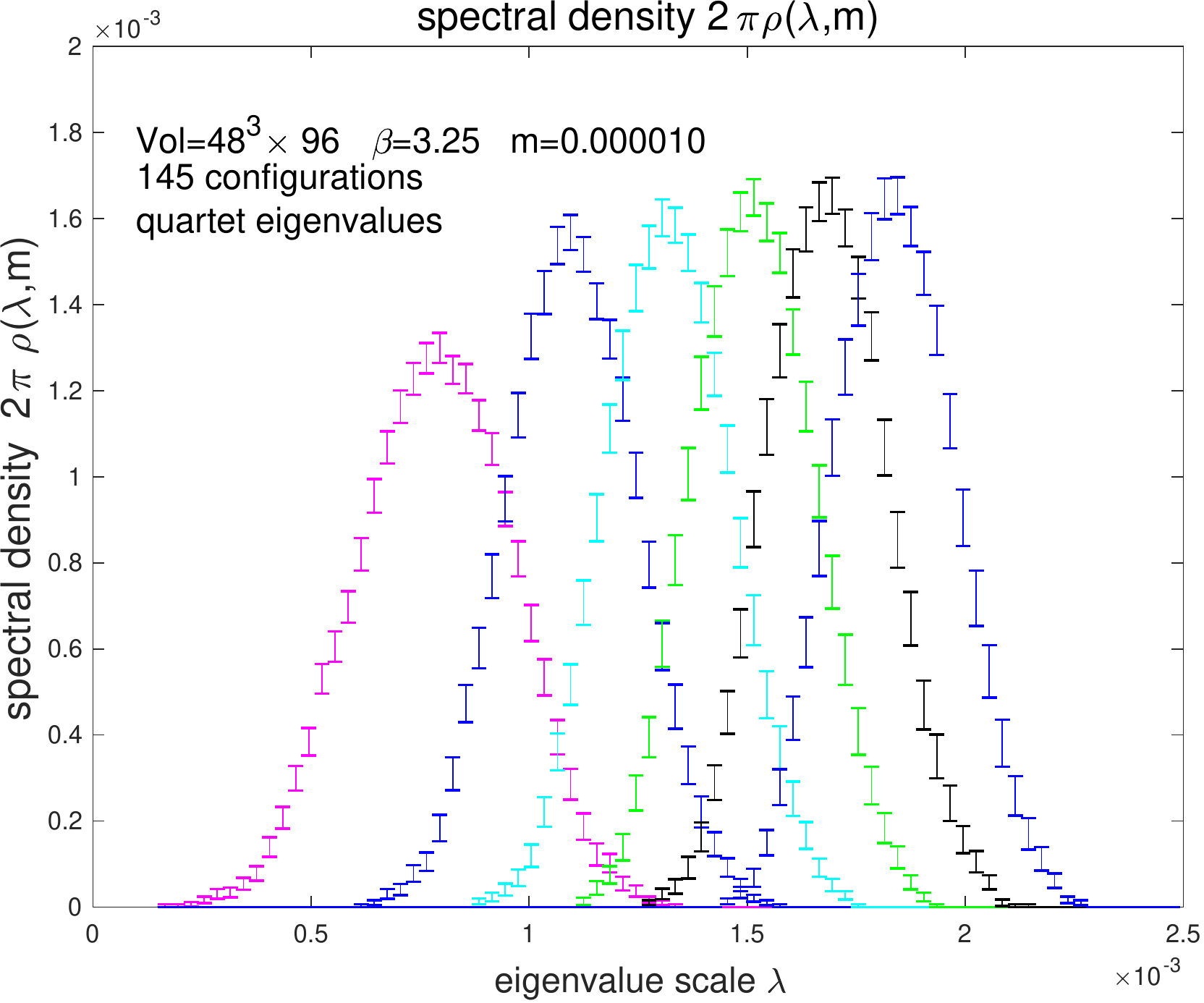}
		\end{tabular}
	\end{center}
	\vskip -0.3in
	\caption{\footnotesize The spectral density of the full spectrum in the sextet model is shown on the left panel.  
    Quartet averages of the lowest 24 eigenvalues are shown on the right panel for input into RMT analysis of the $\epsilon$-regime.
   The shape and distribution of the lowest eigenvalues favors $\chi SB$ close to the chiral limit.}
	\label{fig:RMT}
\end{figure}

In the $\epsilon$-regime, close to the chiral limit where the pion correlation length far exceeds the 
linear size of the finite volume, the EFT Lagrangian of Eq.(\ref{eq:EFT}) simplifies to
\begin{equation}
{\cal L_\epsilon} = \frac{1}{2}\partial_{\mu} \chi \partial_{\mu} \chi\,-\,V_d(\chi)  +
\frac{m^2_\pi f^2_\pi}{4}\big(\frac{\chi}{f_d}\big)^y ~ {\rm tr}\big[\Sigma_0 + \Sigma_0^\dagger\big].\label{eq:EFT1}
\end{equation}
In Eq.(\ref{eq:EFT1}) the coupling of the dilaton to the $\Sigma_0$ zero mode of the pion field is represented by the $\chi(x)$  field and can be treated 
by systematic expansion. In the strict $m\rightarrow 0$ chiral limit the pions become  decoupled from the dilaton field. The challenge of this 
approach is to get close enough to the chiral limit at extremely small fermion masses. 
In fact, from careful studies of the lowest eigenvalues of the Dirac operator we determined that this limit 
would be feasible in  large volume simulations at extremely small $m$ values. The feasibility was demonstrated by decreasing the fermion mass $m$ two orders of magnitude, down to $m=0.000010$ at the sextet gauge coupling $\beta=3.20$  with an estimated  
inverse pion mass of $M^{-1}_\pi \approx 125$ in the equivalent infinite volume p-regime. The simulation results from   
$64^4$ and $48^3\times 96$ lattice volumes with $M_\pi\cdot L < 1$  at $m=0.000010$ are shown 
in Fig.~\ref{fig:RMT}.
\vskip -0.15in
In the limit of asymmetric aspect ratio $L/L_t \rightarrow 0$ at fixed spatial size $L$ we cross over to the $\delta$-regime 
where in the chiral limit the dilaton EFT is further simplified for rotator analysis of the pion and the (de)coupling of the dilaton field, 
\vskip -0.1in
\begin{equation}
{\cal L_\delta} = \frac{1}{2}\partial_{\mu} \chi \partial_{\mu} \chi\,-\,V(\chi) + \frac{f^2_\pi}{4}\big(\frac{\chi}{f_d}\big)^2 
~{\rm tr}\big[\partial_{t}\Sigma_0~\partial_{t}\Sigma_0^\dagger\big].\label{eq:EFT2}
\end{equation}
\vskip -0.1in
The time derivative $\partial_t\Sigma_0$ of the zero three-momentum component of the pion field controls the coupled rotator dynamics 
on the SU(2) group manifold. 
Further analysis of Eqs.(\ref{eq:EFT1},\ref{eq:EFT2}) remains outside the scope of this report.
\vskip 0.1in
\noindent{\bf 6.  Conclusions} 
\vskip -0.02in
Based on the hypothesis of dilaton EFT description, tantalizing test results were obtained from the analysis of the sextet model with particularly interesting physical parameters for the $V_d$ form of the dilaton potential. It is important to note that the dilaton description of the light scalar from broken scale invariance does not follow from the $\beta$-function based walking behavior. Conformal symmetry breaking is not necessarily coupled to walking and requires better theoretical understanding. In addition, extended statistical analysis will be required for the full implementation of the implicit Maximum Likelihood method to assess the sensitivity and quality of our fitting procedure to different forms of the dilaton potential. Ratios of physical parameters depend on the $V_d$  or $V_\sigma$ form of the dilaton potential at fixed lattice spacing, calling for precision studies when the cutoff is varied  in large volumes and close to the chiral limit. The $\epsilon$-regime offers new opportunities, perhaps with direct determination of the effective dilaton potential from methods we developed and tested earlier in Yukawa theories of scalar fields and fermions~\cite{Shen:1988ma,Fodor:2007fn}. It is also an important open question, if the application of the dilaton EFT survives other tests of fermion mass deformations, like $\chi PT$  effects in the chiral condensate, or the renormalized gauge coupling on the gradient flow. 
\vskip 0.1in
\noindent{\bf Acknowledgments}
\vskip -0.02in
We acknowledge support by the DOE under grant DE-SC0009919, by the NSF under grant 1620845,  and by the Deutsche Forschungsgemeinschaft grant SFB-TR 55. Computational resources were provided by the DOE INCITE program on the ALCF BG/Q platform, by USQCD at Fermilab, by the University of Wuppertal, and by the Juelich Supercomputing Center on Juqueen. 
\vskip -0.5in

\bibliographystyle{JHEP}
\bibliography{jkPoS2018}

\end{document}